\pdfoutput = 1
\documentclass[journal]{IEEEtran}
\usepackage{bbm}

\usepackage{epsfig,subfigure}
\usepackage{amssymb,amsmath,amsthm}
\usepackage{enumerate}
\usepackage{latexsym}
\usepackage{booktabs}
\usepackage{multirow} 
\usepackage{array} 
\usepackage{cite}
\usepackage{comment}
\usepackage{todonotes}
\usepackage{authblk}
\usepackage[algoruled]{algorithm2e}
\usepackage{url,verbatim,algorithmic}
\usepackage{tabularx} 
\usepackage{makecell} 
\usepackage{hhline}
\usepackage{graphicx}
\usepackage{epstopdf}
\DeclareGraphicsExtensions{.eps}
\graphicspath{{../}}

\usepackage{amsfonts}
\usepackage{amssymb}
\renewcommand{\vec}[1]{\boldsymbol{#1}}
\newcommand{\vx}{\vec{x}}
\newcommand{\vtx}{\tilde{\vec{x}}}
\newcommand{\vy}{\vec{y}}

\usepackage{bbding}
\usepackage{pifont}
\usepackage{wasysym}
\newcommand{\cX}{\mathcal{X}}

\newtheorem{thm}{Theorem}
\newtheorem{rmk}{Remark}

\newcommand{\ignore}[1]{{}}

\begin{document}
\title{Timely Detection and Mitigation of Stealthy DDoS Attacks via IoT Networks}

\author{Keval Doshi, 
Yasin Yilmaz, and Suleyman Uludag
\thanks{This work was supported in part by the U.S. National Science Foundation under the Grant CNS-1737598 and in part by the SCEEE-17-03 Grant.}
\thanks{
K. Doshi and Y.Yilmaz are with the Electrical Engineering Department, University of South Florida, Tampa, FL USA (e-mail: kevaldoshi@mail.usf.edu, yasiny@usf.edu). }
\thanks{S. Uludag, is with the Department of Computer Science, University of Michigan - Flint, MI USA (e-mail: uludag@umich.edu).}
}

\maketitle

\begin{abstract}
Internet of Things (IoT) networks consist of sensors, actuators, mobile and wearable devices that can connect to the Internet. With billions of such devices already in the market which have significant vulnerabilities, there is a dangerous threat to the Internet services and also some cyber-physical systems that are also connected to the Internet. Specifically, due to their existing vulnerabilities IoT devices are susceptible to being compromised and being part of a new type of stealthy Distributed Denial of Service (DDoS) attack, called Mongolian DDoS, which is characterized by its widely distributed nature and small attack size from each source. This study proposes a novel anomaly-based Intrusion Detection System (IDS) that is capable of timely detecting and mitigating this emerging type of DDoS attacks. The proposed IDS's capability of detecting and mitigating stealthy DDoS attacks with even very low attack size per source is demonstrated through numerical and testbed experiments. 

\end{abstract}

\vspace{-2mm}
\section{Introduction}
\label{s:intro}

The emergence of Internet of Things (IoT) has been one of the most significant technological advances of the last decade\cite{intro}. With the development of various miniaturized embedded systems and many web services along with cloud computing, it is virtually possible to make any isolated system to communicate with another machine. Moreover, the increased capabilities of new System on Chip (SoC) devices, and a drastic reduction in their sizes have led to an exponential increase in the number of devices that communicate through the Internet. With the number of IoT devices in use today already in a few billions, and with an exponential growth, the amount of data generated and transmitted is also witnessing a proportional increase. This has made the IoT paradigm a prime target for a legion of attackers, hackers, cybercriminals and occasionally governments \cite{intro}. 

Unfortunately, the security of IoT devices is not able to keep up with the hardware development and now more and more vulnerabilities are detected on a regular basis leading to security threats and privacy concerns \cite{intro}. 
For example, such compromised devices can be utilized to perform Distributed Denial of Service (DDoS) attacks.
DDoS is a type of cyber-attack in which the perpetuator attacks an online service typically by flooding traffic using a large number of sources. 
Volumetric attacks as the name suggests are characterized by enormous amount of traffic, 
and they normally do not require a large amount of traffic to be generated by the hackers themselves causing it to be the simplest and most common type of DDoS attack \cite{ddos_types}. In this paper, we consider this type of DDoS attacks, especially the stealthy ones, which are challenging to detect and mitigate due to their widely-distributed nature and low-rate anomalous traffic from each source which can easily bypass traditional filters (i.e., stealth attacks). In stealthy DDoS attacks, such as the recent Mongolian DDoS attacks \cite{nexusguard}, although the increase in traffic from each source is small, collectively they are still capable of achieving their goal of disrupting the targeted service due to being widely distributed.
Although a number of practical solutions have been deployed against DDoS, many problems still exist \cite{bertino2017botnets}, especially due to the new genre of DDoS attacks through IoT devices. 
\vspace{-2mm}

\subsection{DDoS via IoT}
\label{s:intro-iot}

There has been a sharp increase in the number of IoT devices with an estimated number of 8.4B devices in 2017 which is expected to reach 20B by 2020 \cite{Smart_Grid}. According to a study by Gartner, a high percentage of new businesses and systems will include an IoT component by 2020\cite{proliferation}. The convenience provided by IoT technologies has led to a wide-scale deployment of a variety of Internet-connected sensors such as thermostats, security cameras, smart lights among many others. Unfortunately, the rapid spread of IoT also brings about a proliferation of security risks. Even though IoT is evolving at an expeditious pace, it is still very much in its inception stage. Hence, at this stage there is a significant risk that hacked IoT devices can be used for nefarious purposes such as being used as a part of a botnet to launch DDoS attacks \cite{bertino2017botnets}.

Currently, IoT network security faces four major challenges:
\begin{itemize}
\item[(C1)] {\it Minimally invasive mitigation:} Because of the distributed nature of recent DDoS attacks, it is very difficult to detect the attacking devices. However, the attack should be mitigated with minimal interruption of services to benign users who want to legitimately use the services under attack. 
\item[(C2)] {\it High dimensionality:} Considering the large number of devices in a typical IoT network, and the abundant data generated by those devices, computationally efficient solutions that can achieve effective network monitoring, i.e., joint monitoring of devices, are required.
\item[(C3)] {\it Unknown attack patterns:} Since there is a wide range of vulnerabilities that attackers can exploit, and new attack techniques are continuously developed by attackers, the predictability of attack patterns is quite low compared to the traditional Internet security. Hence, conventional signature-based detection techniques, as well as parametric probabilistic models are not feasible.
\item[(C4)] {\it Timely detection and mitigation:} Due to the highly interconnected IoT ecosystem including the Internet and critical infrastructure such as Smart Grid, and the potential disastrous effects of cyberattacks, timely detection and mitigation of attacks is crucial. 
\end{itemize}

 We state some of the real-world examples to scrutinize the damage that can be caused by cyberattacks via IoT.

\begin{enumerate}

\item The Mirai botnet, that was launched in 2016, caused one of the most prolific series of attacks in the DDoS history \cite{kolias2017ddos,antonakakis}. This particular botnet infected numerous IoT devices (primarily older routers and IP cameras), and reached data rates higher than 600Gbps. Through flooding the DNS provider Dyn, the Mirai botnet took down many popular websites such as Etsy, GitHub, Netflix, Shopify, SoundCloud, Spotify, and Twitter. Mirai took advantage of devices running out-of-date firmware, and relied on the fact that most users do not change the default usernames/passwords on their devices. The fact that its source code has been released on a hacker forum (and now it is available online \cite{Mirai_Code}) facilitated its derivations. 

\item In November 2016, hackers shut down the heating of two buildings in the city of Lappeenranta, Finland. This was another DDoS attack, and in this case, the attack was specifically targeted towards an attribute of smart home. The attackers managed to cause the heating controllers to continually reboot the system in a loop so that the heaters never worked. The attack was significant because the temperatures are well below freezing at that time of the year, and such a scenario can be life threatening.

\item In early 2017, Verizon Wireless released a report that included an unnamed university that experienced an attack from more than 5,000 IoT devices, such as vending machines and smart light bulbs. 

\end{enumerate}
These examples portray the severity of the situation if such botnets are acquired by sophisticated hackers and employed against critical infrastructures such as Nuclear Plants, Smart Grids, etc. 
\vspace{-3mm}

\subsection{Related Works}

DDoS attacks via IoT networks are relatively less addressed compared to other security issues in the IoT enviornment. However, it is recently attracting considerable interest, e.g., \cite{bertino2017botnets},\cite{kolias2017ddos},\cite{antonakakis}. In \cite{lopez2019network,yusof2019systematic,shtern2014towards,cambiaso2012taxonomy}, a wide range of vulnerabilities because of which conventional signature-based detection techniques fail, are discussed. In \cite{Kashi}, authors propose a solution to UDP flood attack in an IoT environment using 6LoWPAN and IEEE  802.15.4. However, it has high overheads and complex architectures and components which do not suit an IoT environment \cite{PrevWork}. An agent-based DDoS mitigation approach is proposed in \cite{Sonar}. The authors propose a two-part algorithm in which the attack detection part has been performed in the border router. 
A similar entropy-based solution is proposed in \cite{Entropy}, but with the requirement that the packet contents should be detectable. They also do not consider scenarios in which the entropy does not change but the number of packets do.  Xiang et al. \cite{Informetric} proposes an information metric approach to quantify the differences between legitimate and attack network traffic by assuming that the legitimate traffic follows a Gaussian distribution, whereas the attack traffic follows a Poisson distribution. 
Machine learning algorithms are also gaining attention as recent anomaly detection research shows promise \cite{chandola2009anomaly}. Doshi et al. \cite{doshi2018} presents the performance of popular machine learning algorithms such as SVM, k-nearest-neighbors, neural networks etc. in detecting malicious traffic. However, they require training data for malicious traffic (supervised anomaly detection), and extract features which are specific to certain IoT devices without considering other devices that might be present in the network such as laptops or smartphones. In \cite{unsuper}, Nomm et al. proposes using feature selection with popular anomaly detection techniques such as one-class SVM to detect IoT botnet attacks. They use three different approaches for extracting useful features and provides their results over the N-BaIoT dataset. Kurt et al. \cite{kurt2018} proposes an online anomaly detection algorithm based on dimensionality reduction, that is capable of detecting anomalies in high dimensional settings. They also present their results using the N-BaIoT dataset. Meidan et al. \cite{meidan2018n} proposes using deep autoencoders for detecting DDoS attacks at the network level. They achieve small false positive rates by training a deep autoencoder for each individual device in the network, which might not scale well to large networks with many devices. They also use a window-based majority voting scheme to detect attacks, which is not well suited for quick detection. 

\vspace{-3mm}
\subsection {Contributions}
\vspace{-1mm}

In this paper, we propose a practical anomaly-based detection and mitigation technique for IoT-based DDoS attacks, especially the challenging \emph{stealthy DDoS attacks with data rate increase per device as low as 10\%}, which is significantly lower than the considered rates in the literature, and can easily bypass most of the existing approaches. Specifically, the proposed technique is based on a statistical anomaly detection algorithm called Online Discrepancy Test (ODIT) that mitigates the attack with minimal interruption of regular service; scales well to large systems; does not rely on presumed baseline and attack patterns; and achieves quick and accurate detection and mitigation thanks to its sequential nature. The major contributions of this paper are as follows: 
\begin{itemize}
    \item 
    A novel detection and mitigation technique for stealthy DDoS attacks is proposed, and its time and space complexity is analyzed; 
    \item 
    Asymptotic optimality of the proposed detector is proven in the minimax sense as the training data size grows; 
    \item 
    Solution to a dynamic scenario in which the number of devices in the network changes is provided; 
    \item 
    A comprehensive performance evaluation is provided using a testbed implementation, the N-BaIoT dataset, and simulations.
\end{itemize}
 
We first present the problem formulation in Sec. \ref{s:prob}, then provide the proposed IDS in Sec. \ref{s:odit}, experimental and testbed results in Sec. \ref{s:sim}, discuss the limitations in Sec. \ref{s:limit}, and finally conclude the paper in Sec. \ref{s:conc}.

\section{Problem Formulation and Background}
\label{s:prob}

\subsection{System Model}

\begin{figure}[th]
\centering
\includegraphics[width=.45\textwidth]{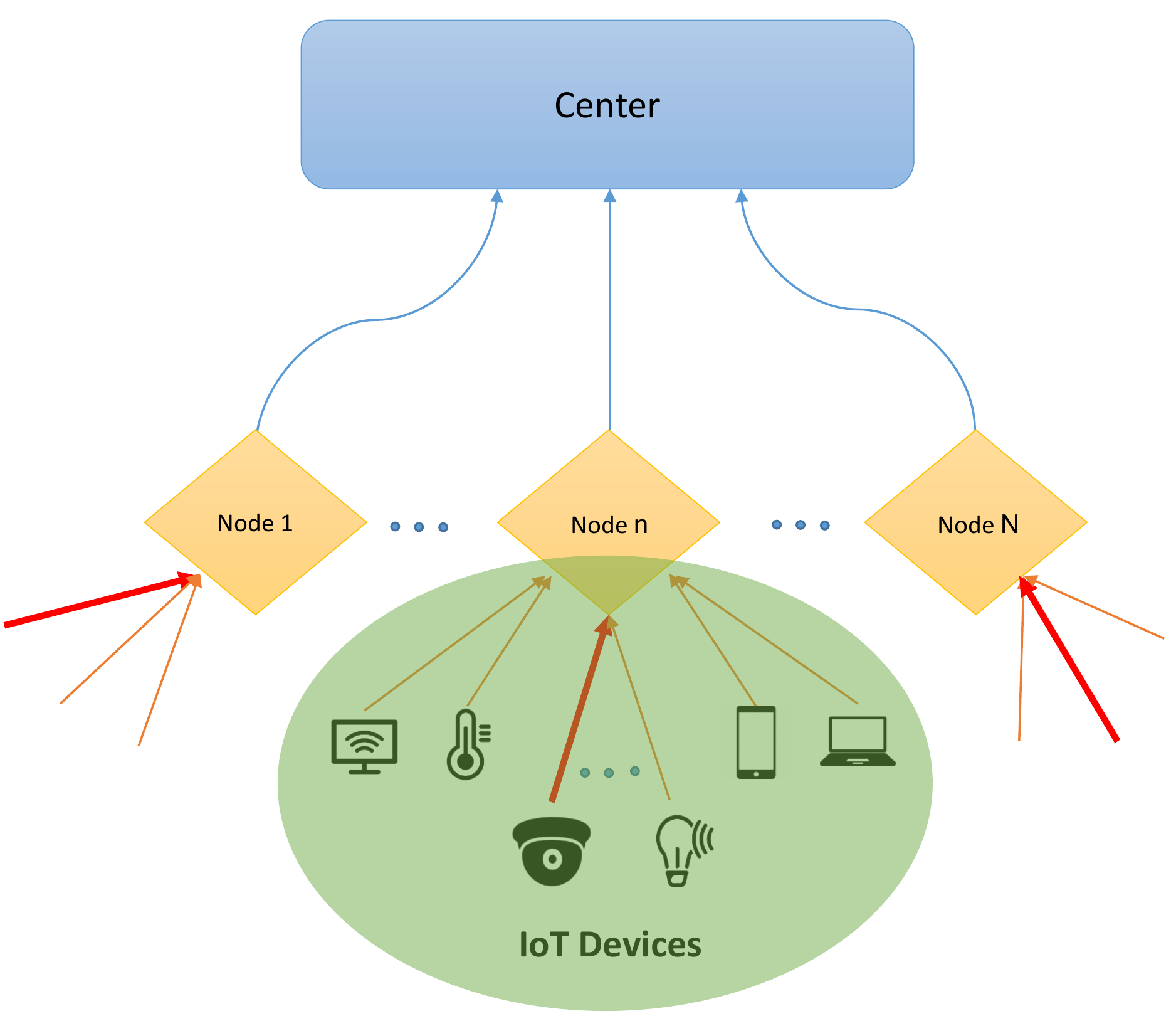}
\vspace{-2mm}
\caption{System model consisting of IoT devices such as thermostat, CCTV, light bulb, smartphone, etc. In the threat model, the bold arrows imply an increased packet rate.}
\label{f:no_anomaly}
\vspace{-5mm}
\end{figure}

In the considered architecture (Fig. \ref{f:no_anomaly}), each IoT device sends its data to the node connected to it. Nodes direct the data traffic to a center, such as a web server, data center or utility center. The architecture is scalable in such a way that a node may represent a smart home consisting of tens of devices or a smart building/neighborhood access point consisting of thousands of devices.

Each device typically has different data communication characteristics. In particular, the data content is typically different (for example, a thermostat would have considerably smaller packet size as compared to a CCTV camera), and the communication protocol used might be different (such as TCP, UDP or HTTP). Also, for the same device the data rates might differ significantly based on the location or type of connection, e.g., a laptop connected via a fiber optic cable might send 1000 packets per second whereas the same device would be sending 10 packets per second on a slower connection. Even the active communication frequency varies considerably, e.g., devices like printers update its status once a minute, whereas CCTV cameras send data every second. In this work, our only assumption is that they perform a packet-based data communication. 
\vspace{1mm}

\subsection{Threat Model}

We consider a volumetric DDoS attack scenario in which data rates (packet/sec.) from a number of devices increase at some point in time (see Fig. \ref{f:no_anomaly}). Particularly, we consider a threat model in which some IoT devices are compromised and start to send more than usual number of data packets. We do not assume further attack specifications such as knowledge on how devices are compromised (e.g., through a vulnerability in the firmware, spoofing attack, man-in-the-middle attack, use of default password), the attack magnitude (i.e., percentage of increase) and duration, and whether the data content changes or not. It is not tractable to mitigate such attacks at the center since accurate identification of all attacking devices is not tractable due to the highly distributed nature of the attacks. Moreover, due to the low-rate nature of the attacks, it is very difficult to detect them locally at the nodes. We propose a general Intrusion Detection System (IDS) that can run locally and is capable of detecting and mitigating an attack even when it is not possible to inspect the data content, which is a prerequisite for many IDS algorithms, e.g., \cite{Entropy}, \cite{Limm}.

Although the standard volumetric attacks that are studied in the literature include high increase in the data rate of a device, with the increasing number of compromised IoT devices low-rate stealthy DDoS attacks (e.g., with a 20\% increase per device) started to become threatening \cite{nexusguard}. Due to the proliferation of IoT, cyber-criminals can launch widely-distributed and highly-effective stealthy DDoS attacks, that can bypass conventional filters and IDSs. Hence, in this paper we study DDoS attacks with data increase rates as low as 20\% per device. There are existing works which consider low-rate DDoS attacks (e.g., \cite{Informetric}), nevertheless the considered increase rates are still significantly higher than what we consider in this paper.

As a result of such widely-distributed  DDoS attacks (e.g., the almost uniform distribution of attack traffic in Mirai \cite{incap}), it is not tractable to have a single global solution running at the server end, and thus we propose in the next section a local IDS that runs at each node. Such local solutions also facilitate accurate mitigation. 

\ignore{

\begin{table}
\centering
\begin{tabular}{ |p{3cm}|p{3cm}|  }

 \hline
 \multicolumn{2}{|c|}{Country List} \\
 \hline
 Country Name& \% Mirai Botnet IPs\\
 \hline
Vietnam&	12.8\%\\
Brazil&		11.8\%\\
United States&10.9\%\\
China&		8.8\%\\
Mexico&	8.4\%\\
South Korea	&6.2\%\\
Taiwan&	4.9\%\\
Russia	&	4.0\%\\
Romania&	2.3\%\\
Colombia&	1.5\%\\
\hline

\end{tabular}
\vspace{3mm}
\caption{Countries of origin of Mirai DDoS Attacks}
\label{table:1}
\end{table}

\begin{figure}[t]
\centering
\includegraphics[width=.47\textwidth]{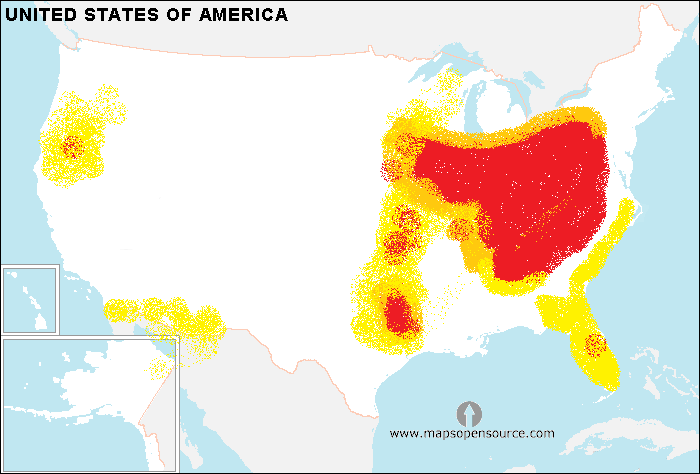}
\caption{Distribution of attack locations across United States in Mirai Botnet \cite{Wiki}.}
\label{f:outage}
\end{figure}

}
\section{Proposed Anomaly-Based IDS }
\label{s:odit}

\begin{figure}[ht]
\centering
\includegraphics[width=.5\textwidth]{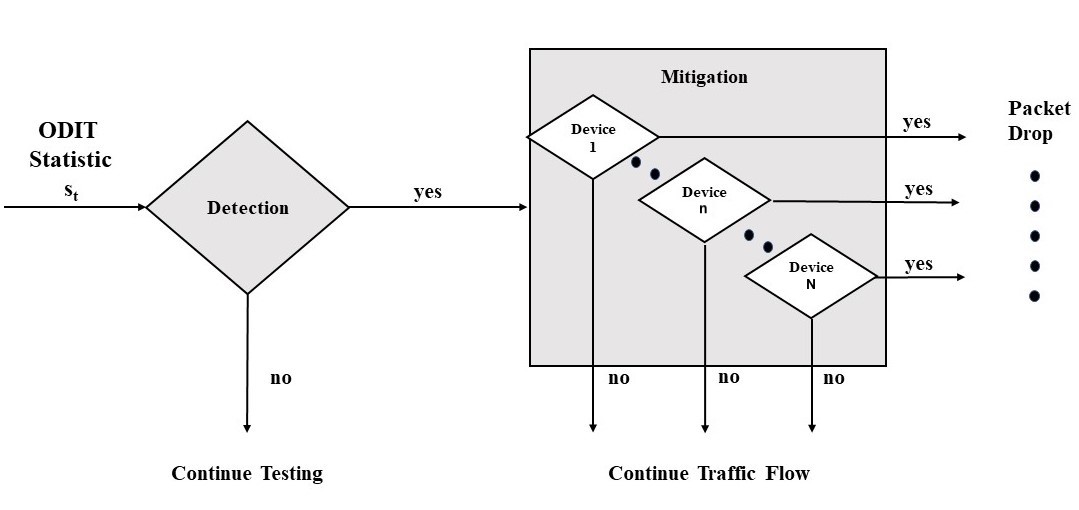}
\vspace{-3mm}
\caption{Proposed Intrusion Detection and Mitigation System }
\label{f:IDS}
\vspace{-5mm}
\end{figure}

In this section, we present our detection and mitigation strategy for the proposed IDS. As shown in Fig. \ref{f:IDS}, we detect an attack based on the cooperative test statistic, and once an attack is detected, we monitor each device individually to identify the attacking devices (see Sections \ref{s:coop} and \ref{sec:miti}). We also analyze the computational complexity and a practical scenario in which the number of devices in the network are dynamic in nature.

\subsection{Proposed Detection Strategy}
\label{s:algorithm}
The heterogeneous nature of an IoT network makes parametric anomaly detection approaches for DDoS detection less effective since they assume probabilistic models for nominal and anomalous conditions. In practice, it is difficult to know/estimate the anomalous and even the nominal probability distributions. Hence, parametric anomaly-based IDSs, as well as many conventional signature-based IDSs are not feasible in addressing stealthy DDoS attacks through IoT. Recently, an online and non-parametric detector called the Online Discrepancy Test (ODIT) was proposed for detecting persistent and abrupt anomalies \cite{ODIT}. Thanks to its nonparametric operation, ODIT does not need to know baseline or anomalous distributions beforehand, hence can address the challenge (C3) stated in Section \ref{s:intro-iot}. ODIT is a sequential method which accumulates evidence in time, and makes a decision at each time based on the accumulated evidence so far,  instead of making a hard decision based on a single data point. This sequential nature of ODIT is tailored for timely detection, thus it is able to address the challenge (C4). Moreover, ODIT can handle monitoring large number of devices together (see Sections \ref{s:complex} and \ref{s:sim}), which addresses the challenge (C2). 

\begin{figure}[t]
\centering
\includegraphics[width=.5\textwidth]{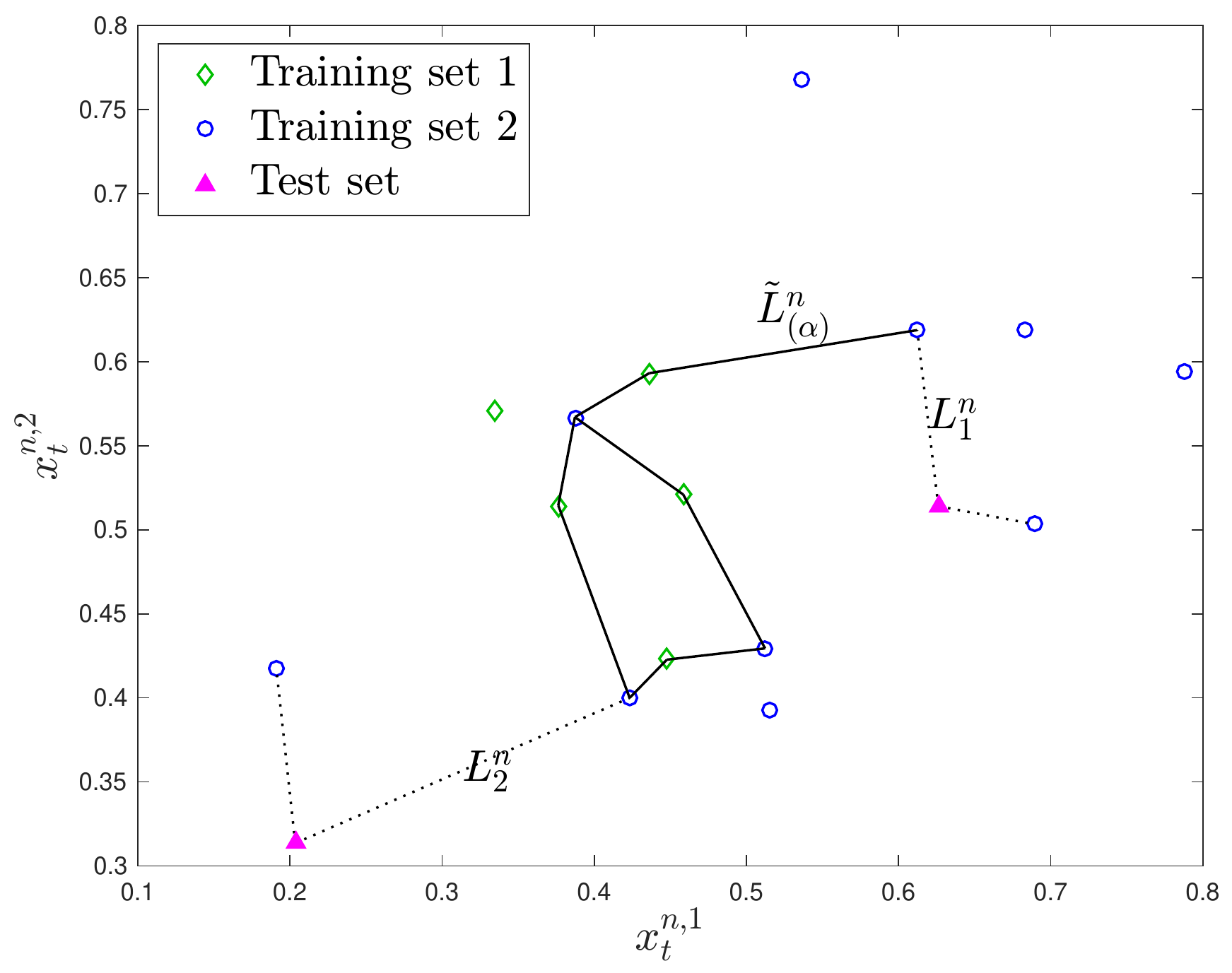}
\vspace{-3mm}
\caption{ODIT procedure with $M_1=5$, $M_2=10$, $\alpha=0.2$, $k=2$. $D_1^n=d[\log L_1^n-\log \tilde{L}_{(\alpha)}^n]$ and $D_2^n=d[\log L_2^n-\log \tilde{L}_{(\alpha)}^n]$ are used as in \eqref{e:Dt} for online anomaly detection (see also Fig. \ref{f:odit_stat}). Test points are from the same nominal distribution as training points, which is a two-dimensional Gaussian with independent components, $0.5$ mean, and $0.1$ standard deviation.}
\label{f:graph}
\vspace{-5mm}
\end{figure}


In this work, we propose a novel modification for ODIT, and prove that this modified version, as the training data size grows, asymptotically becomes the Cumulative Sum (CUSUM) test, which is the optimum sequential change detection algorithm in the minimax sense. CUSUM is a parametric test which assumes both the nominal and the anomalous distributions are completely known \cite{Bass}. 

We next show the procedure for the proposed ODIT-based IDS for a node $n$, which observes a $d$-dimensional normalized data vector $\vx_t^n \in [0,1]^d$, where $d$ is the number of devices, at each time $t$. Here, in the context of DDoS attack detection, $\vx_t^n$ denotes the number of packets received from the $d$ devices in the network at time $t$ and normalized by the corresponding maximum number of packets for each device.  Normalization of each dimension of $\vx_t^n$ into $[0,1]$ is performed to deal with the typical heterogeneity in the data communication characteristics of IoT devices.Then, in Section \ref{s:coop}, we show how to achieve cooperation among nodes.

\textbf{Training:} Given an attack-free training dataset $\mathcal{X}^n_M=\{\vtx^n_1,\ldots,\vtx^n_M\}$ which represents the baseline (i.e., nominal) operation, we begin our training procedure.
\begin{enumerate}

\item Randomly split $\cX^n_M$ into two subsets $\mathcal{X}^n_{M_1}$ and $\mathcal{X}^n_{M_2}$ with $M_1$ and $M_2$ points, where $M_1+M_2=M$, for computational efficiency, as in the bipartite GEM algorithm\cite{Srichanran}. 
\item For each point $\vtx^n_i$ in $\mathcal{X}^n_{M_1}$ find the $k$th-nearest-neighbor ($k$NN) distance $\tilde{L}_i^n$ with respect to the points in $\mathcal{X}^n_{M_2}$.
\item For a significance level $\alpha$, e.g., $0.05$, store the $(1-\alpha)$th percentile $\tilde{L}_{(\alpha)}^n$ of $k$NN distances $\{\tilde{L}_{1}^n,\ldots,\tilde{L}_{M_1}^n\}$ to use as a baseline statistic for computing the anomaly evidence of test instances.
\end{enumerate}

The training procedure is illustrated in Fig. \ref{f:graph}, where the training set consists of $M=15$ points, which is then randomly split into two sets of $M_1=5$ points (denoted by green) and $M_2=10$ points (denoted by blue). In this example, for each point in the first set, we find the second-nearest-neighbor ($k=2$) distance with respect to the second set. The largest $k$NN distance ($\alpha=0.2$) among the points in the first set is used as the baseline statistic $\tilde{L}_{(\alpha)}^n$.

\textbf{Testing:} When a new data $x_t^n$ is observed at time $t$, 
\begin{enumerate}
\item Compute
\begin{equation}
\label{e:Dt}
D_t^n = d[\log L_t^n - \log \tilde{L}_{(\alpha)}^n]
\end{equation}
where $L_t^n$ is the $k$NN distance of the new data point $x_t^n$ with respect to the points in $\mathcal{X}^n_{M_2}$, and $\tilde{L}_{(\alpha)}^n$ is obtained in the training.

\item Treating the statistic $D_t^n$ as a positive/negative evidence for anomaly accumulate it over time as in CUSUM:
\begin{equation}
\label{e:stat}
s_t^n = \max\{s_{t-1}^n + D_t^n,0\}, s_0^n=0,
\end{equation}
where $D_t^n$ is given in \eqref{e:Dt}. 

\item Decide to continue taking a new data point $x_{t+1}^n$ if the accumulated evidence $s_t^n$ is not sufficient for raising an attack alarm, and stop and raise an alarm at the first time $s_t^n \ge h$, i.e., at time
\begin{equation}
\label{e:stop}
T_n = \min\{t: s_t^n \ge h_n\},
\end{equation}
where $h_n>0$ is a predetermined threshold. 
\end{enumerate}

\begin{figure}[t]
\centering
\includegraphics[width=.5\textwidth]{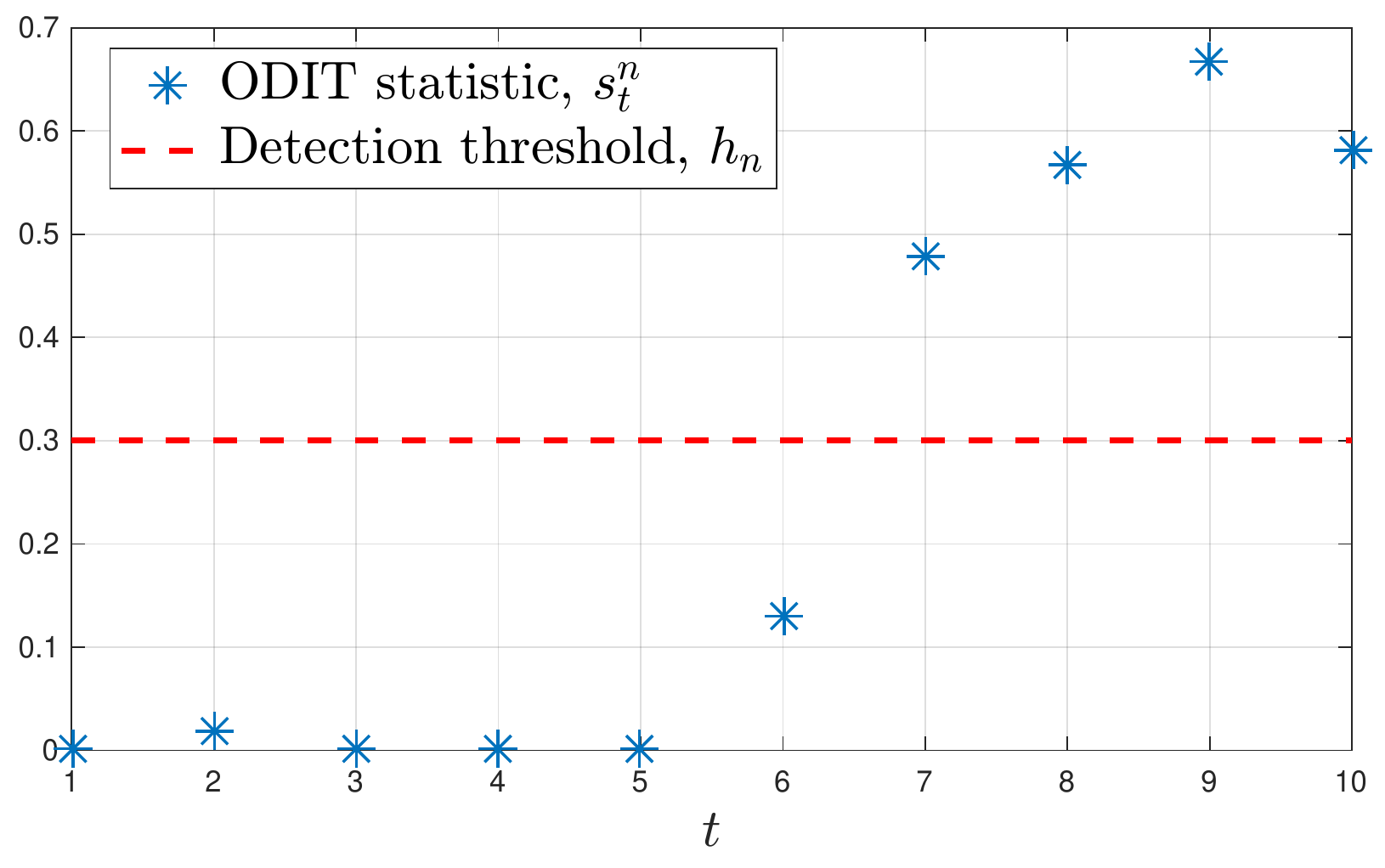}
\vspace{-3mm}
\caption{ODIT statistic and decision procedure using the setup in Fig. \ref{f:graph} and anomalous test points from uniform distribution over $[0,1]$. Anomaly starts at $t=6$, and detected at $t=7$ with the shown threshold.}
\label{f:odit_stat}
\vspace{-2mm}
\end{figure}

As compared to anomaly evidence $D_t=L_t-\tilde{L}_{(\alpha)}$ presented in the original ODIT algorithm \cite{ODIT}, we use the form given in Eq. \ref{e:stat}. This modification enables an asymptotic optimality proof, which is presented in Theorem \ref{thm:RAPID}. The intuition behind this modification is to explicitly show the analogy between the attack evidence $D_t$ and log-likelihood ratio. 

\begin{thm}
\label{thm:RAPID}
When the nominal distribution $f_0(\vx_t^n)$ is finite and 
continuous, and the attack distribution $f_1(\vx_t^n)$ is a uniform distribution whose support includes $\vx_t^n$, as the training set grows, the anomaly evidence $D_t^n$ converges in probability to the log-likelihood ratio,
\begin{equation}
D_t^n \overset{p}{\to} \log \frac{f_1(\vx_t^n)}{f_0(\vx_t^n)}
~~~ \text{as}~~~ M_2 \to \infty,
\end{equation}
i.e., the proposed ODIT detector converges to CUSUM, which is minimax optimum in minimizing expected detection delay while satisfying a false alarm constraint.
\end{thm}
\begin{IEEEproof}
See the Appendix.
\end{IEEEproof}

Since the proposed detector does not train on anomalous data or assume any model for anomaly, the uniform distribution condition on $f_1$ for asymptotic optimality is expected.

\textbf{Parameter Selection:} The detection threshold $h_n$ manifests a trade-off between minimizing the detection delay and minimizing the false alarm rate, as can be seen in Fig. \ref{f:odit_stat}. Particularly, smaller threshold facilitates early detection, but also increases the probability of false alarm. In practice, $h_n$ can be chosen to satisfy a given false alarm rate. 
The number of neighbors $k$ also affects the trade-off between early detection and small false alarm rate. Smaller $k$ would result in being more sensitive to anomaly, hence supports earlier detection, but at the same time it causes to be more prone to the false alarms due to nominal outliers. Larger $k$ would result in vice versa.
The choice for $M_1$ and $M_2$ is typically skewed towards $M_2$, i.e., $M_2>M_1$, since $M_2$ determines the degree of resemblance between $k$NN distance likelihood under the nominal case, as explained in Theorem \ref{thm:RAPID}. The significance level $\alpha$ is an intermediate parameter whose effect can be compensated by the threshold $h$. As a rule of thumb, a small $\alpha$ value, such as $0.05$, should be first selected, and then $h_n$ should be set to satisfy a desired false alarm rate. 

\begin{rmk}
A training set that is free of anomaly can be obtained through either human supervision or through an isolated secure system. We should emphasize here the difference between anomaly and outlier. The training set may contain outliers that are generated under no-attack conditions. Outliers correspond to ``tail events" that can occur under nominal settings with low probability. Although outliers are rare under normal operations, they can still exist in the training set, and their natural existence does not harm the regular operation of the proposed detector. On the other hand, anomaly is a change in the system behavior, i.e., in the probability distribution of the generated data. 
In other words, anomaly can be defined as the existence of ``persistent outliers", as opposed to the sporadic nominal outliers.
\end{rmk}


\subsection{Cooperative Operation}
\label{s:coop}

Multiple nodes running the proposed IDS given in Eq. \eqref{e:Dt}--\eqref{e:stop} can cooperate for earlier detection and mitigation of attacks by leveraging the hierarchical structure shown in Fig. \ref{f:no_anomaly}. Following the cooperative CUSUM with independence assumption in \cite{mei2010efficient} we propose a cooperative detector which sums the local statistics computed at the nodes to obtain a global statistic $s_t = \sum_{n=1}^N s_t^n$. That is, at each time $t$, each node $n$ updates its local statistic $s_t^n$ using \eqref{e:stat} and transmits it to the center, which sums them to obtain the global detection statistic $s_t$. Then, the center decides whether or not there is an attack similarly to \eqref{e:stop}, i.e., raises an alarm at time $T=\min\{t:s_t\ge h\}$.
This cooperative detector can detect attacks earlier than the single-node ODIT detector thanks to the spatial diversity, i.e., accumulated attack evidences from multiple nodes. Note that the statistics from nodes without any attacked device typically take small values close to zero, but never become negative according to \eqref{e:stat}. Thus, they do not negatively contribute to the global statistic $s_t$, and consequently do not cause extra delay in detection.

The cooperative scheme through summing local statistics in a hierarchical architecture enables the proposed detector to easily scale to arbitrarily large networks. The optimal statistical detection would normally require multivariate analysis for all the devices in the entire network. However, due to the abundance of IoT devices this is not feasible; and more importantly due to the natural hierarchical structure in IoT networks such a large-scale multivariate analysis is unnecessary. Specifically, IoT devices are grouped under nodes such as smart home routers, and devices under different nodes can be typically modeled independently under no-attack conditions. Although an attack will normally correlate them, it is reasonable to relax that constraint since a practical attack strikes devices asynchronously, i.e., each of the attacked devices has different attack onset times. This relaxation is critical for CUSUM to be applicable to any multi-dimensional setting since it is not tractable to estimate the set of attacked devices to perform multivariate analysis \cite{mei2010efficient}.
After detection, the mitigation procedure given in the next subsection can be applied at each node.

\subsection{Mitigation Strategy}
\label{sec:miti}

Timely detection of a DDoS attack is a necessary, but not a sufficient condition for ensuring the security of a system. We need a mitigation strategy that is capable of stopping the DDoS attack by identifying the attacking IoT devices, and then blocking the traffic originating from those devices. 

We perform an in-depth analysis to determine which IoT devices are causing the increase. We begin by examining the cooperative statistic $s_t$ calculated in Sec. \ref{s:coop} and determining which nodes are causing the increase. Once the attacking nodes are identified, we examine every dimension, which represent an IoT device connected to the node, of the distance $L_t^n$, calculated in Sec. \ref{s:algorithm}. 
$(L_t^n)^2$ corresponds to the squared Euclidean norm of the $d$-dimensional distance vector $\vy_t^n=\vx_t^n-\vtx_{(k)}^n$ whose $j$th entry $y_t^{n,j}$ is the distance of the data from device $j$ at time $t$ to the $j$th dimension of $k$th nearest neighbors, i.e., $(L_t^n)^2 = \|\vy_t^n\|_2^2$. If $y_t^{n,j}$ comes out large, then it contributes to a large $L_t^n$ towards an alarm, which provides an evidence that the device is under attack. Hence, after an alarm is raised at time $T$, we 
\begin{enumerate}
\item first determine the time instance $\tau$ when the test statistic $s_t$ started to increase since the last time it was zero ($\tau=6$ in Fig. \ref{f:odit_stat}), which can be seen as an estimate for the attack onset time,
\item then, compute the average statistic
\begin{equation}
\label{eq:nmiti}
\bar{s}_n  =  \frac{1}{T-\tau+1} \sum\limits_{t=\tau}^{T} s_t^n
\end{equation}
for each node $n$, and compare it with a threshold $\theta_1$ to determine the attacking nodes, i.e., node $n$ has attacked devices if $\bar{s}_n \ge \theta_1$.
\item then, for each node identified as attacking, compute the average distance
\begin{equation}
\label{eq:miti}
\bar{y}_{n,j}  =  \frac{1}{T-\tau+1} \sum\limits_{t=\tau}^{T} y_t^{n,j}
\end{equation}
for each device $j$ under it, and compare it with a threshold $\theta_2$ to decide as attacking or not, i.e., device $j$ is identified as attacking if $\bar{y}_{n,j} \ge \theta_2$.
\end{enumerate}

As usual the selection of threshold $\theta_1$ and $\theta_2$ controls a balance between the False Positive Rate (FPR) and True Positive Rate (TPR). As shown in Fig. \ref{f:roc_iot}, the proposed mitigation technique achieves high TPR for almost all FPR even in the challenging attack scenario investigated in Section \ref{s:sim}.
We should note that the procedure in Eq. \eqref{eq:miti} requires some memory to store the most recent distance values for all dimensions local to the node performing the procedure.

Combining the detection and mitigation strategies, the proposed IDS technique is summarized in Algorithm \ref{alg}.

\begin{algorithm}                      
\caption{Proposed detection \& mitigation algorithm}          
\label{alg}                           
\begin{algorithmic}[1]                    
    \STATE {\bf Initialize:} $s_0=0$, $t=0$
    \FOR{$n = 1,\ldots,N$}
        \STATE Partition training set into $\cX^n_{M_1}$ and $\cX^n_{M_2}$
        \STATE Determine $\tilde{L}_{(\alpha)}^n$
    \ENDFOR 
    \WHILE{$s_t < h$}
        \STATE $t \gets t+1$ 
        \STATE Get new data $\{x_t^n\}$ and compute $\{D_t^n\}$ as in \eqref{e:Dt}
        \STATE $s_t^n = \max\{s_t^n+D_t^n,0\}$   
        \STATE $s_t = \sum_{n=1}^N s_t^n$.
    \ENDWHILE
       \STATE Declare attack at $T=t$
\FOR{$n = 1,\ldots,N$}
\STATE Compute $\bar{s}_n$ as in \eqref{eq:nmiti}
	\IF{$\bar{s}_n \ge \theta_1$}
	\FOR{$j=1,\ldots,d$}
	\STATE Compute $\bar{y}_{n,j}$ as in \eqref{eq:miti}
	\IF{$\bar{y}_{n,j} \ge \theta_2$}
		\STATE Block traffic from device $j$
	\ENDIF
\ENDFOR
	\ENDIF
\ENDFOR

\end{algorithmic}
\end{algorithm}

\subsection{Computational Complexity}
\label{s:complex}

The following theorem shows that the proposed algorithm can scale well to large systems.
\begin{thm}
The online time complexity and space (i.e., memory or storage) complexity of Algorithm \ref{alg} linearly scales with $M_2$, the number of points in the second training set, and $d$, the number of devices, i.e., $O(M_2 d)$. The offline training time complexity is $O(M_1 M_2 d)$.
\label{thm:comp}
\end{thm}
\begin{IEEEproof}
See the Appendix. 
\end{IEEEproof}

\begin{rmk}
There are efficient ways of finding (approximate) $k$ nearest neighbors that scale even better to high-dimensional systems. For instance the method proposed in \cite{muja2014scalable} has a time complexity of $O(\lambda d)$ in online testing where $\lambda$ is the maximum number of points to examine for finding $k$ nearest neighbors. $\lambda$ can be chosen much smaller than $M_2$ at the expense of decreasing the accuracy of $k$NN approximation. 
Hence, a balance between approximation quality and computational complexity should be sought while choosing $\lambda$.
Consequently, using a fast $k$NN method instead of straightforward computation the real-time operation capability and/or the scalability of the proposed method can be significantly enhanced.
\end{rmk}

\subsection{Dynamic Environments}


A major challenge for any anomaly-based intrusion detection system is adaptability to dynamic environments. This means that the system should be adaptable to changes in the environment, while still recognizing abnormal activities. 
In a dynamic IoT network such as a university or a shopping mall, the number of devices may frequently change based on the number of people, time of the day, day of the year etc.
With varying number of devices over time the challenge for the proposed IDS is computing the $k$NN distance under varying number of data dimensions. 

A key observation here is that data rates are specific to applications rather than devices. For instance, video streaming has certain data rates regardless of the streaming device. Hence, considering a list of applications that are used in the network, such as web browsing, music and video streaming, we can deal with the changing number of dimensions. Specifically, we first collect training data for the extreme scenario with maximum number of devices running each application simultaneously. During online testing, at each time we modify the training set by ignoring the unused dimensions for each application, and compute $L_t^n$. 

\begin{figure}[t]
\centering
\includegraphics[width=.5\textwidth]{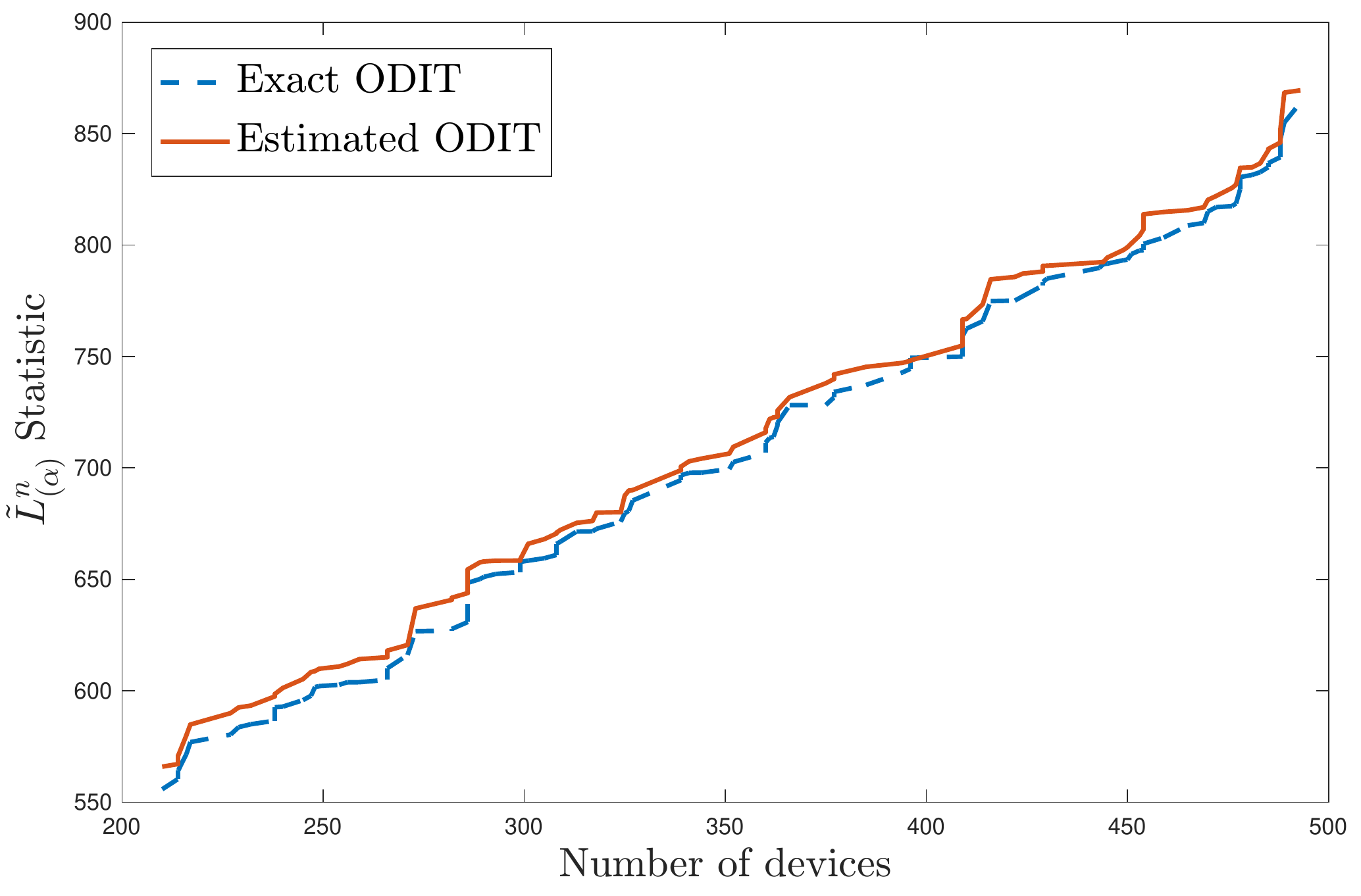}
\vspace{-3mm}
\caption{Comparison of the estimated $\tilde{L}_{(\alpha)}^n$ statistic to the computed one for different combinations with increasing number of devices.}
\label{f:estimated_odit}
\vspace{-5mm}
\end{figure}

Since there is a huge number of possible combinations for the number of devices running each application, performing the expensive training procedure (see Theorem \ref{thm:comp}) for each such combination to compute the baseline statistic $\tilde{L}_{(\alpha)}^n$ is not feasible. 
Hence, we propose to build a function approximator for $\tilde{L}_{(\alpha)}^n$. 
We collect data for several different combinations, and compute $\tilde{L}_{(\alpha)}^n$ for each such combination. Using the number of devices running each application as input we train a regression model to estimate the $\tilde{L}_{(\alpha)}^n$ value for a given combination. The results for Gaussian process regression is shown in Fig. \ref{f:estimated_odit}. A simpler method (e.g., linear regression) or a more sophisticated method (e.g., deep neural network) can be used for the regression model. In Fig. \ref{f:estimated_odit}, we compare the estimated $\tilde{L}_{(\alpha)}^n$ statistic to the computed one for different combinations with increasing number of devices. We see that the estimated statistic closely matches that of exact ODIT, which is infeasible to compute for all combinations. Furthermore, the baseline statistic $\tilde{L}_{(\alpha)}^n$ depends on the selected significance level $\alpha$, which is a design parameter. Our simulations show that a small mismatch between the estimated and computed $\tilde{L}_{(\alpha)}^n$ values is not critical for the algorithm's performance.

\section{Experimental Results}
\label{s:sim}


In this section we evaluate the performance of the proposed IDS using real data, an IoT testbed, and simulations.

\subsection{N-BaIoT Dataset}
\begin{table*}[thb]
  \caption{
Overview of the N-BaIoT dataset properties, botnet infections, and considered attack types}
  \centering
\resizebox{\textwidth }{!}
{
    \begin{tabular}{c||ll||cc||ccccc||ccccc}
    \hhline{===============}
    & \multicolumn{2}{c}{\textbf{Dataset Properties}} & 
    \multicolumn{2}{c}{\textbf{Botnet Infections}} &
    \multicolumn{5}{c}{\textbf{Considered Bashlite Attacks}} & 
    \multicolumn{5}{c}{\textbf{Considered Mirai Attacks}} \\
    \hline
     \begin{tabular}[t]{@{}c@{}}\textbf{Device}\\\textbf{ID}\end{tabular} 
& \textbf{Device Make and Model}         
& \textbf{Device Type}            
& \textbf{Mirai}
& \textbf{BASHLITE} 
& \begin{tabular}[t]{@{}c@{}}\textbf{Combo}\end{tabular} 
& \begin{tabular}[t]{@{}c@{}}\textbf{Junk}\end{tabular}
& \begin{tabular}[t]{@{}c@{}}\textbf{Scan}\end{tabular}
& \begin{tabular}[t]{@{}c@{}}\textbf{TCP}\end{tabular} 
& \begin{tabular}[t]{@{}c@{}}\textbf{UDP}\end{tabular} 
& \begin{tabular}[t]{@{}c@{}}\textbf{ACK}\end{tabular} 
& \begin{tabular}[t]{@{}c@{}}\textbf{Scan}\end{tabular}
& \begin{tabular}[t]{@{}c@{}}\textbf{SYN}\end{tabular}
& \begin{tabular}[t]{@{}c@{}}\textbf{UDP}\end{tabular}
& \begin{tabular}[t]{@{}c@{}}\textbf{UDP Plain}\end{tabular} \\
    \hhline{===============}
    1   & Danmini                & Doorbell      & \Checkmark & \Checkmark  & \Checkmark   & \Checkmark & \Checkmark & - & - & \Checkmark & \Checkmark & \Checkmark & - & \Checkmark  \\
    2   & Ennio                  & Doorbell       & x & \Checkmark & \Checkmark   & \Checkmark & \Checkmark & - & - & - & - & - & - & -  \\
    3   & Ecobee                 & Thermostat     & \Checkmark & \Checkmark & \Checkmark   & \Checkmark & \Checkmark & - & - & \Checkmark & \Checkmark & \Checkmark & - & \Checkmark  \\
    4   & Philips B120N/10     & Baby Monitor   & \Checkmark & \Checkmark & \Checkmark   & \Checkmark & \Checkmark & - & - & \Checkmark & \Checkmark & \Checkmark & - & \Checkmark  \\
    5   & Provision PT-737E        & Security Camera & \Checkmark & \Checkmark & \Checkmark   & \Checkmark & \Checkmark & - & - & \Checkmark & \Checkmark & \Checkmark & - & \Checkmark  \\
    6   & Provision PT-838          & Security Camera & \Checkmark & \Checkmark & \Checkmark   & \Checkmark & \Checkmark & - & - & \Checkmark & \Checkmark & \Checkmark & - & \Checkmark  \\
    7   & SimpleHome XCS7-1002-WHT & Security Camera & \Checkmark & \Checkmark & \Checkmark   & \Checkmark & \Checkmark & - & - & \Checkmark & \Checkmark & \Checkmark & - & \Checkmark  \\
    8   & SimpleHome XCS7-1003-WHT & Security Camera & \Checkmark & \Checkmark & \Checkmark   & \Checkmark & \Checkmark & - & - & \Checkmark & \Checkmark & \Checkmark & - & \Checkmark  \\
    9   & Samsung SNH 1011 N	            & Webcam        & x & \Checkmark  & \Checkmark   & \Checkmark & \Checkmark & - & - & - & - & - & - & -  \\
    \hhline{===============}
    \end{tabular}
}    
\label{tab:decrip}
\end{table*}

\begin{figure}[t]
\centering
\includegraphics[width=.4\textwidth]{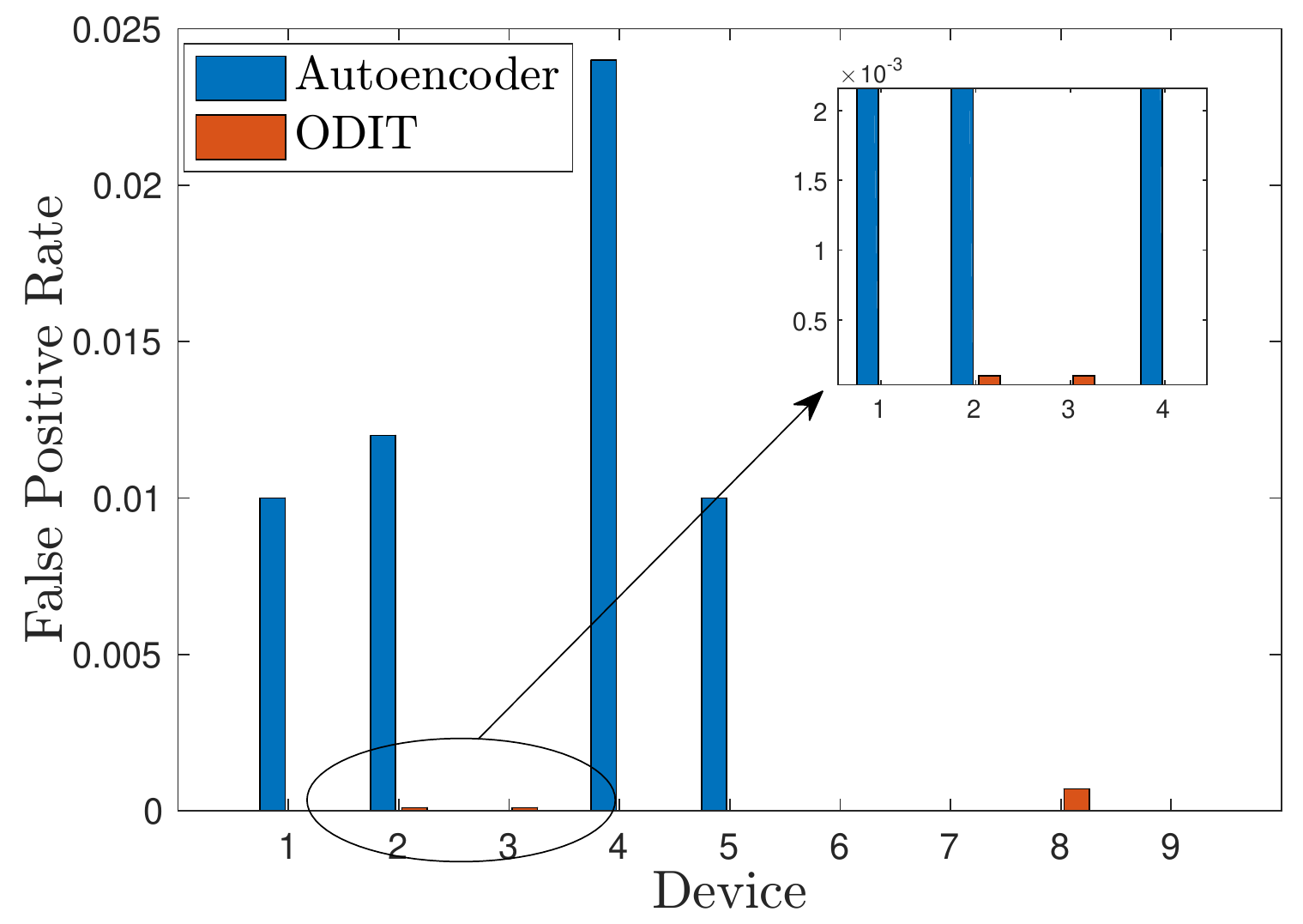}
\includegraphics[width=.37\textwidth]{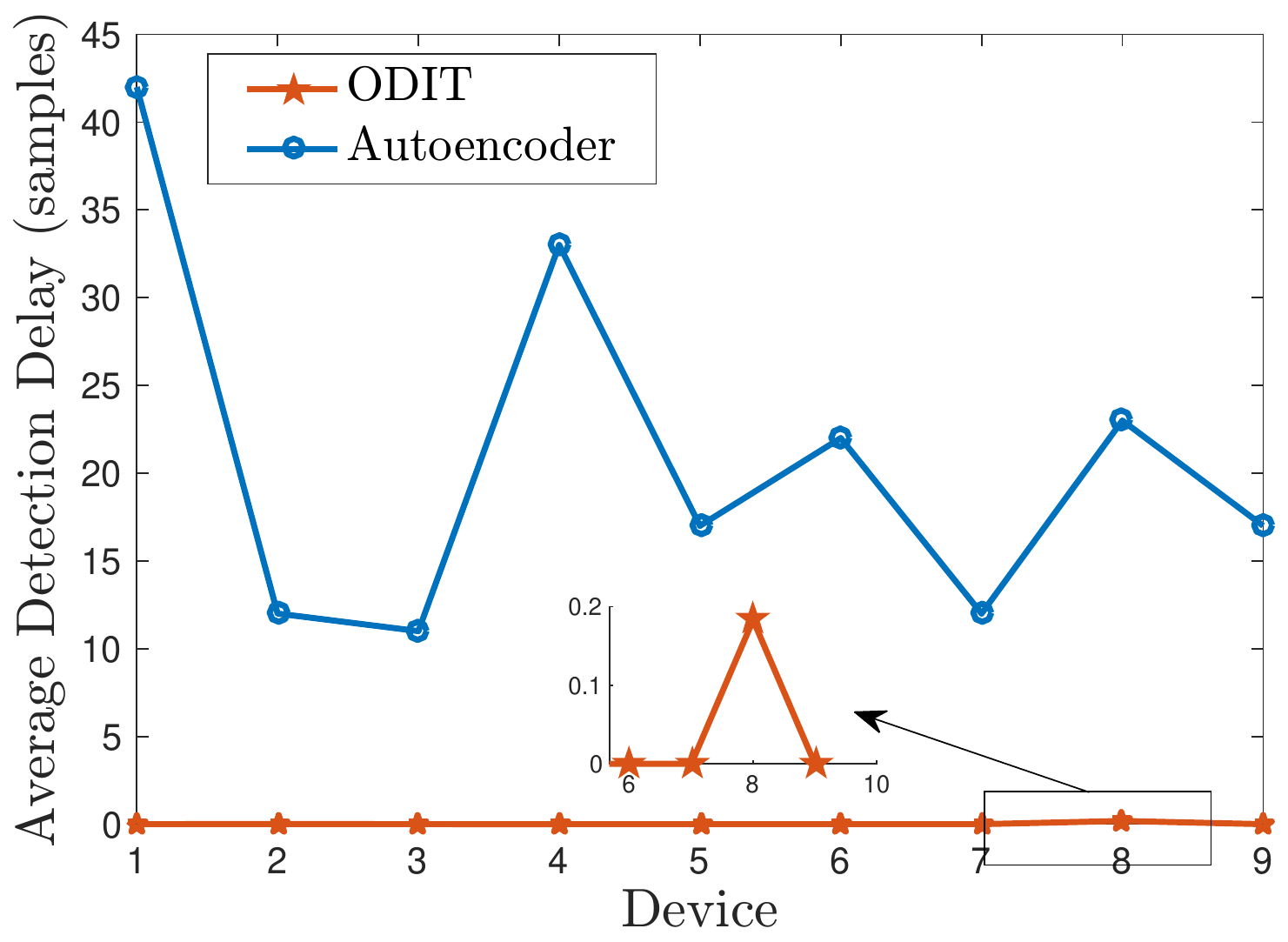}
\vspace{-3mm}
\caption{Comparison between deep autoencoder-based IDS and the proposed ODIT-based IDS in terms of false positive rate (top) and average detection delay (bottom). The x-axis corresponds to the index of the attacked device.}
\label{f:FPR_nb}
\vspace{-2mm}
\end{figure}

We firstly consider the N-BaIoT dataset \cite{meidan2018n} \footnote{This dataset is available at the UCI Machine Learning Repository.}, which contains data from various IoT devices under both nominal and attack conditions. The network consists of of nine devices, namely a thermostat, a baby monitor, a webcam, two doorbells, and four security cameras connected via WiFi. The dataset description, as well as attacks considered in our experiments are presented in Table \ref{tab:decrip}. We do not consider UDP and TCP attacks in our experiments since the provided data does not correlate with typical UDP and TCP attacks.

Our results show that the proposed ODIT-based IDS significantly outperforms the deep autoencoder-based IDS proposed in the N-BaIoT paper \cite{meidan2018n}, and by extension Isolation Forest \cite{isolation}, SVM \cite{svm} and LOF \cite{lof}, which are shown to be outperformed by the autoencoder method. In Fig. \ref{f:FPR_nb}, we present the false positive rate and the average detection delay when each device is under attack. Here, the average detection delay is analogous to the false negative rate as all the misclassified anomalous instances would contribute to the average detection delay. The proposed ODIT-baed method achieves much more accurate and quicker detection than the autoencoder-based method except for device 8. On further analyzing the data, we see that there are a few outliers in the data for device 8, which causes some false alarms. Note that by increasing the decision threshold $h$, we are able to reduce the false positive rate at the cost of a slightly higher detection delay. Conversely, a smaller detection delay can be achieved by setting a lower threshold at the cost of a few more false alarms. 
Thanks to its sequential nature, the ODIT-based IDS detects the attacks right after it occurs while satisfying very small false alarm rates. Whereas, the autoencoder-based IDS applies a majority voting in a moving window for attack detection, thus its detection delay is at least half the window size. The optimum window sizes reported in \cite{meidan2018n} for each device are used for comparisons.

\subsection{Testbed Results}
\label{s:exp}

\begin{figure}[t]
\centering
\includegraphics[width=0.45\textwidth]{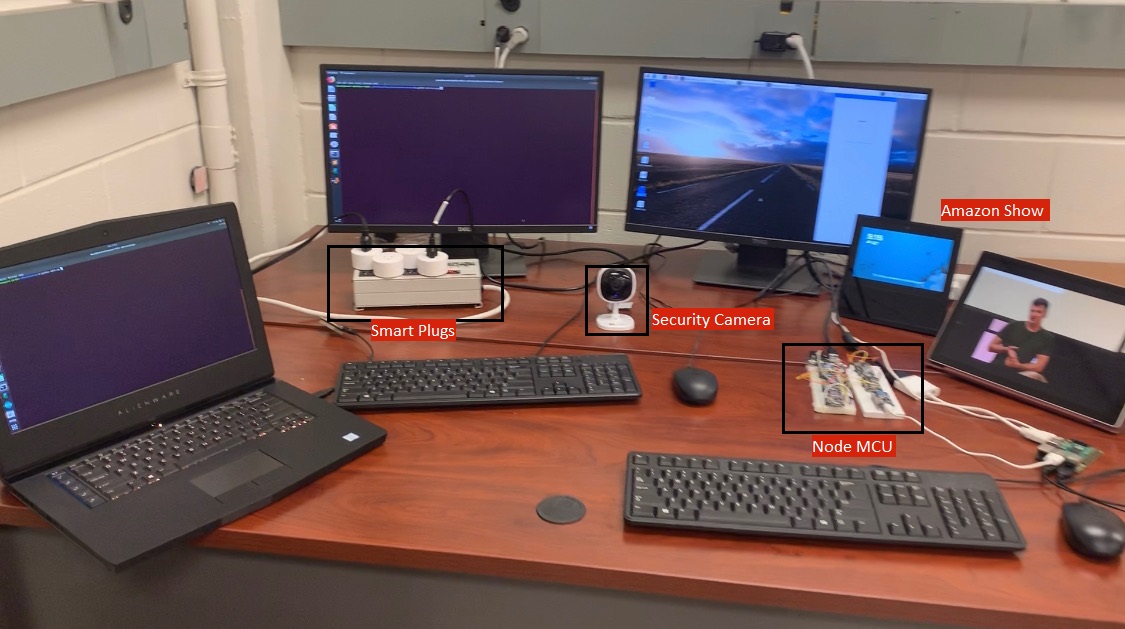}
\caption{IoT testbed consisting of NodeMCUs, smart switches, security camera, Amazon Echo Show, laptop, tablet, and Raspberry Pi.}
\label{f:Testbed}
\vspace{-5mm}
\end{figure}

 \begin{figure}[t]
\centering
\includegraphics[width=.4\textwidth]{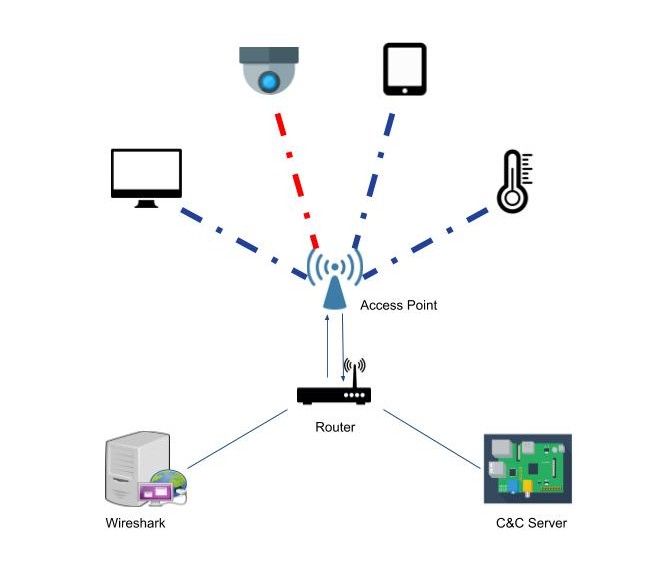}
\vspace{-3mm}
\caption{Setup for hardware implementation.}
\label{f:setup-iot}
\vspace{-5mm}
\end{figure}

\begin{figure}[t]
\centering
\includegraphics[width=.5\textwidth]{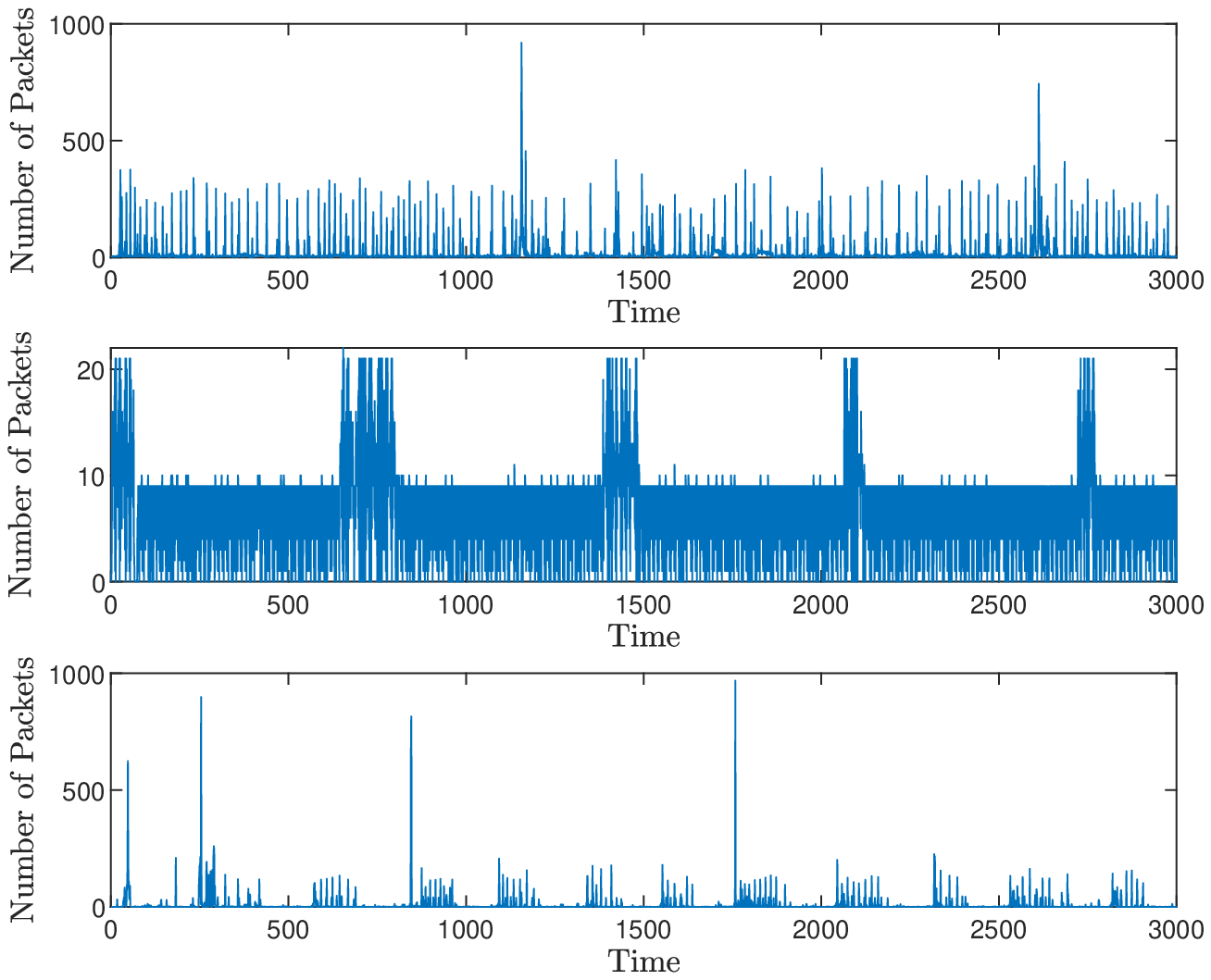}
\vspace{-5mm}
\caption{Time series for number of packets sent by a computer (top), a NodeMCU (middle), and an Amazon Echo Show (bottom).}
\label{f:pac_num}
\vspace{-3mm}
\end{figure}

We next present a hardware implementation of our proposed attack detection and mitigation algorithm. Even though there are several dataset already available, none of them addresses stealthy DDoS attacks. In existing works, even a 300\% increase from the nominal data rate of a device is considered low-rate DDoS, e.g., \cite{Informetric}. However, due to the exponential increase in the number of IoT devices, much lower data rates, e.g., a 30\% increase, might be sufficient for performing effective DDoS attacks, as exemplified by the recent Mongolian DDoS attacks \cite{nexusguard}. Also, most of the datasets seem to concentrate only on vulnerable devices, but in a typical network, there are also devices that cannot be compromised easily, and account for a significant amount of background data. We first present the testbed setup, and then provide the experimental results using the testbed data. 

\textbf{\textit{Testbed Setup:}}
To demonstrate a typical IoT network, we collected the network traffic data from devices that were connected via Wi-Fi to an access point, which is wire connected to a router. For sniffing the network traffic, we performed port mirroring on the router, and recorded the data using Wireshark.
The goal here is to design a setup that is as close to a real life scenario as possible for studying stealthy DDoS attacks. We consider a network consisting of 15 popular IoT devices, namely a laptop computer, a tablet, 7 NodeMCUs, 4 smart switches, an Amazon Echo Show device, and a security camera as shown in Fig. \ref{f:Testbed}. The purpose of using a computer and a tablet is to consider devices which cannot be easily compromised yet they account for a significant amount of traffic passing through the router. The NodeMCUs, which may represent various IoT devices on the market, are configured to update their status on a local server. A Raspberry Pi acts as a command and control (C\&C) server which is used to start and stop attacks. In Fig. \ref{f:setup-iot}, we show the setup for our hardware implementation. A comparison between the number of packets transmitted from a computer, a NodeMCU, and an Amazon Echo Show over a period of time is shown in Fig. \ref{f:pac_num}. It is seen that each device has two major operating states: an active state and an idle state.  However, there is a stark difference between the transmission patterns for each device. 

 \begin{figure}[t]
\centering
\includegraphics[width=.5\textwidth]{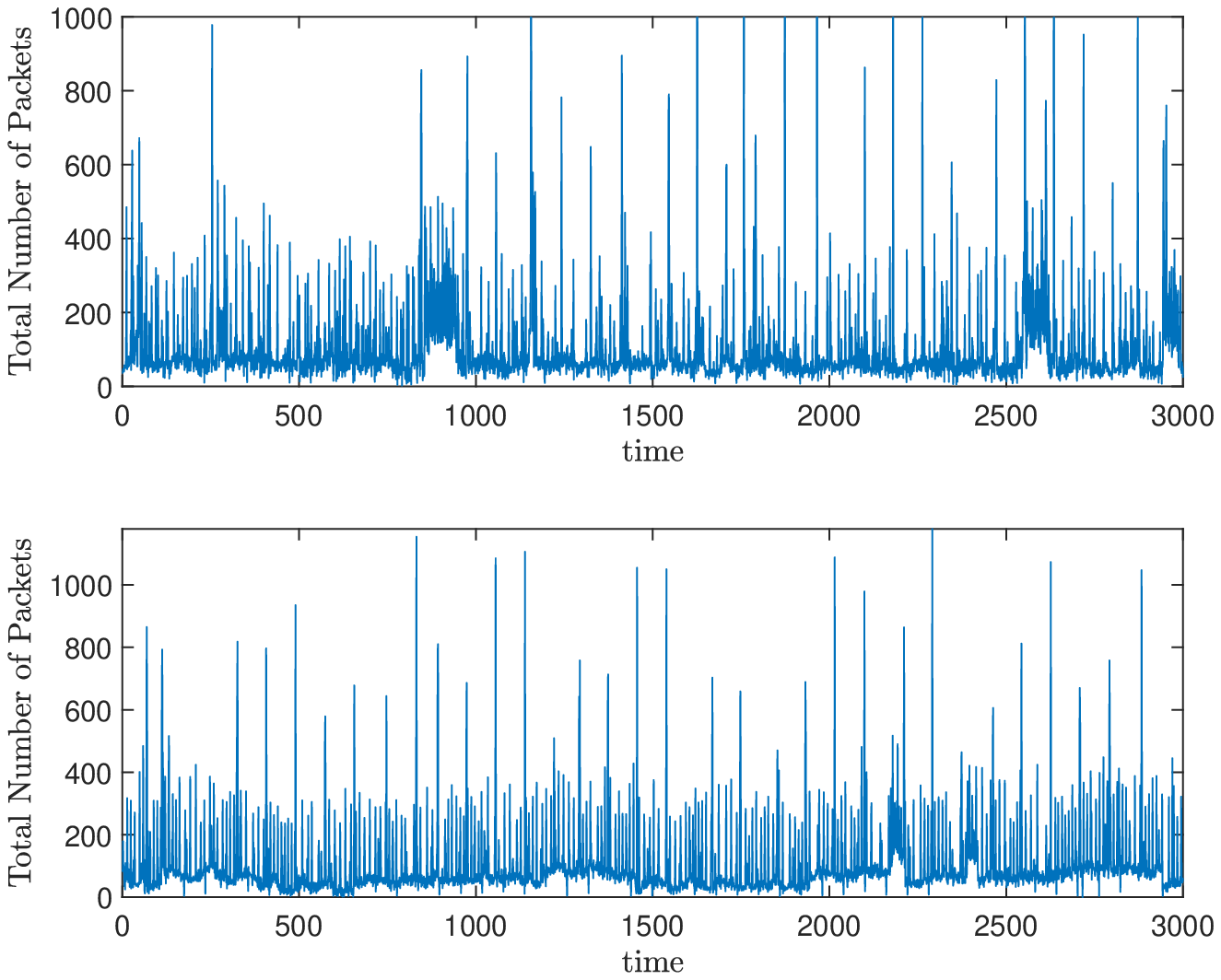}
\vspace{-5mm}
\caption{Time series for total number of packets sent over the entire network without (top) and with the stealthy Http flooding attack (bottom).}
\label{f:tot_num}
\vspace{-3mm}
\end{figure}
 
\begin{figure*}[t]
\centering
\includegraphics[width=.9\textwidth]{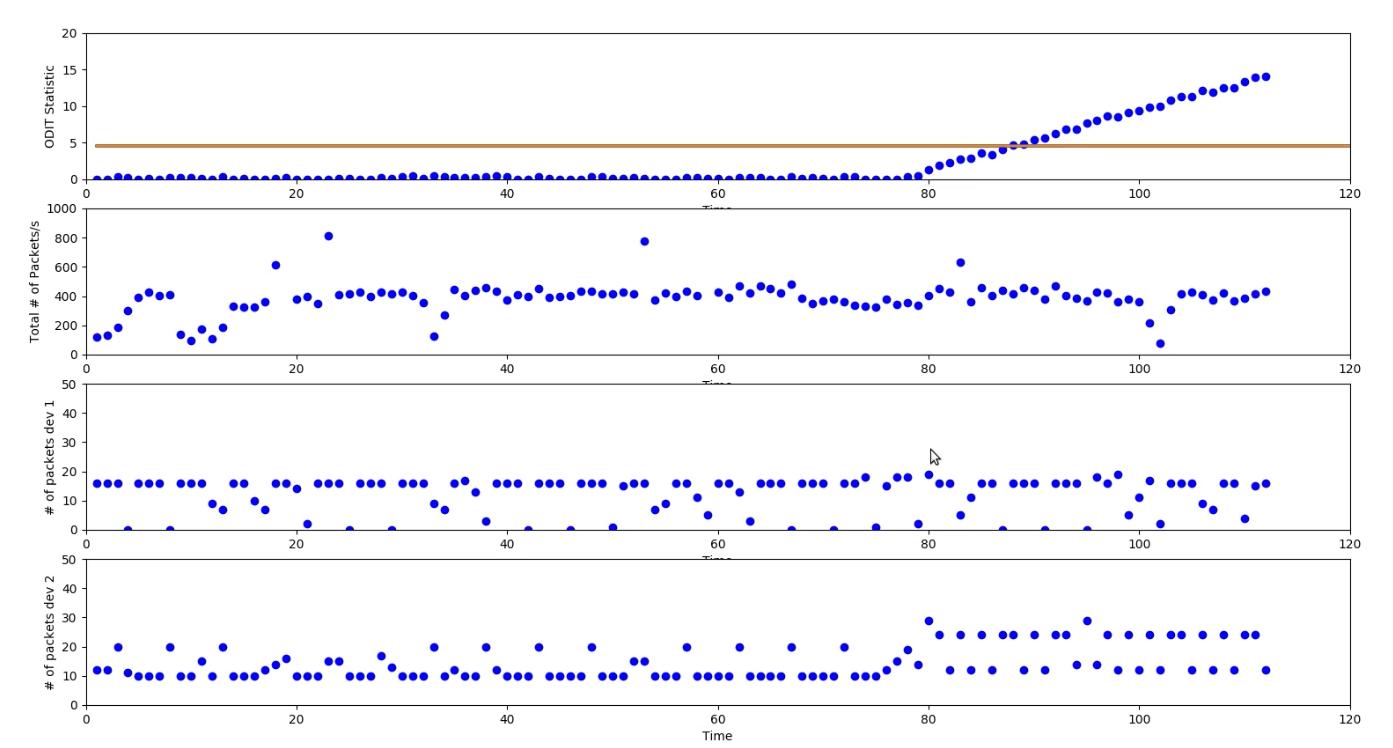}
\vspace{-3mm}
\caption{Live implementation of the proposed IDS in the stealthy Http flooding attack case. At $t=80$ attack starts by increasing the data rate of device 1 by 10\% (third from top) and device 2 by 30\% (bottom). In such a stealthy attack, there is no visible change in the total number of packets in the network (second from top). The detection statistic of the proposed IDS steadily increases right after the attack, and alarms when it crosses the threshold (top).}
\label{f:Live_stat}
\end{figure*}

\begin{table}[t]
	\caption{Testbed Attack Characteristics}
	\centering
	\resizebox{0.5\textwidth}{!}{
		\begin{tabular}{c|c|c|c}
			\hhline{====}	
			\multicolumn{1}{c}{\textbf{Attack Name}}& 
			\multicolumn{1}{c}{\textbf{Attack Type}}& 
			\multicolumn{1}{c}{\textbf{Magnitude}}& 
			\multicolumn{1}{c}{\textbf{CompromisedDevice}} \\
			\hhline{====}
			Http Flooding & Application & Low-Rate & Node MCU \\
			\hline
			ICMP Flooding & Volumetric & Low-Rate & Node MCU \\
			\hline
			Ping of Death & Protocol & Low-Rate & Laptop \\
			\hline
			UDP Flooding & Volumetric & High-Rate & Laptop \\
			\hhline{====}	
		\end{tabular}
	}
	\label{tab:extracted_features}
	\vspace{-3mm}
\end{table}

\begin{figure*}
\centering
\includegraphics[width=1\textwidth]{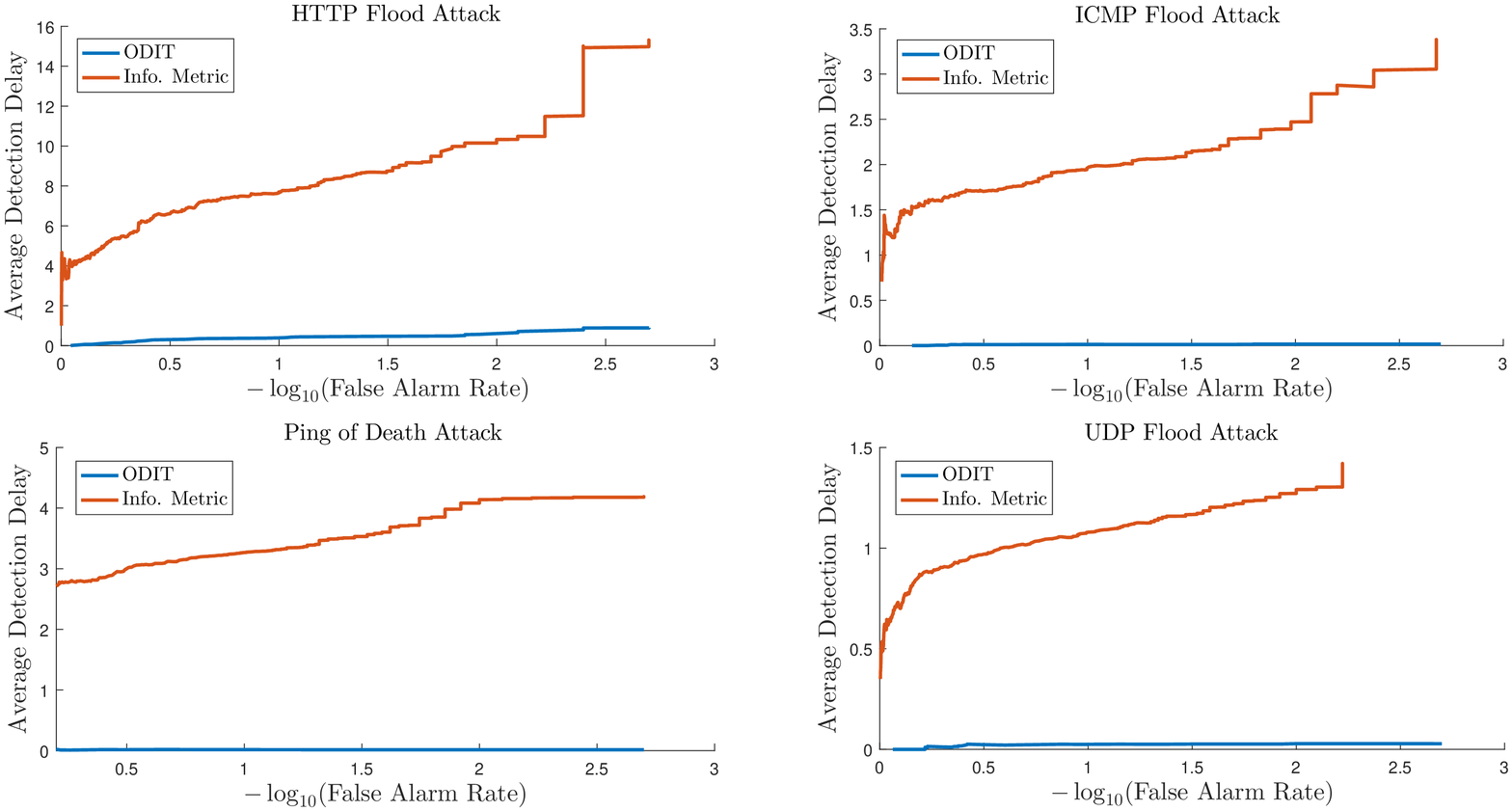}
\vspace{-5mm}
\caption{Comparison of the proposed ODIT-based IDS and the information metric-based IDS in \cite{Informetric} for attacks presented in Table \ref{tab:extracted_features}.}
\label{f:All_attack}
\vspace{-5mm}
\end{figure*}

\textbf{\textit{Attack Models \& Results:}}
We implemented 4 different attacks, as shown in Table \ref{tab:extracted_features}. The data was captured in pcap format by using Wireshark and is publicly available \footnote{\url{https://github.com/kevaldoshi17/DDoSAttackDetection}}. 
In Http flooding, at $t=80$ sec., we slightly increase the mean rates of updating the server for two of the NodeMCUs. Particularly, there is small increase in the number of \textit{Get} requests from two NodeMCUs. To depict that the attack magnitude does not have to be consistent across all devices, we increase the mean packet rate of NodeMCU 1 by 10\% and of NodeMCU 2 by 30\%. In Fig. \ref{f:tot_num}, we show a comparison between the total number of packets sent over the entire network with and without the stealthy Http attack. It is seen that due to the low-rate nature of the attack, there is no visible increase in the total number of packets. Hence, filter-based methods which monitor the total number of packets transmitted in the network, would fail to detect such a stealthy attack. 
In all cases, the input to the proposed IDS is the number of each packet type from each individual device.

In Fig. \ref{f:Live_stat}, we see that from $t=0$ to $t=80$, the ODIT statistic under the Http attack does not increase considerably, but after $t=80$, it steadily increases. This figure is taken from a real-time demonstration which is available online \footnote{\url{https://youtu.be/zQexZgB5AMs}}. By adjusting the threshold we can have a trade-off between the detection delay and number of false alarms. 
In Fig. \ref{f:All_attack}, we compare the proposed ODIT-based IDS with the information metric-based algorithm proposed in \cite{Informetric} in terms of average detection delay under all attack cases. The method in \cite{Informetric} uses an information distance metric based on the generalized (R{\'e}nyi) entropy. It uses a window to compute the information metric on the aggregate traffic at each node, which causes loss in time resolution, and also in early detection ability. 
Since ODIT monitors each packet type from every individual device, it is able to detect the attack with a much smaller detection delay for the same false alarm rates. 

Similarly in the other three attack cases, ODIT quickly and accurately detects the attacks by closely monitoring the data traffic in each type from each device thanks to its multivariate nature. It takes much longer for the information metric IDS to detect the attacks at the same false alarm rate as there is no significant increase in the number of total number of packets. 
Although in ICMP flooding, the data rates of the two NodeMCUs are again increased 10\% and 30\%, this time it is easier for the information metric method to detect since the number of ICMP packets in the network is much smaller than the number of Http packets. 
In the case of ping of death attack, which results in an increase in the number of ICMP packets, the proposed IDS achieves zero detection delay in all trials. 
Finally, in the UDP flood attack, we considered a higher attack rate by increasing the nominal data rate of the laptop by 100\%. In this case, the performance of information metric method improves, but ODIT still outperforms it by detecting the attack under 0.2 second on average for a false alarm rate of 0.01. 

\subsection{Simulation Results}

\begin{table}
\caption{Nominal data model for different IoT devices.}
\centering
\begin{tabular}{ |p{2cm}||p{1.3cm}|p{.8cm}||p{1.3cm}|p{.8cm}|  }
 \hline
 \multicolumn{5}{|c|}{Data Model} \\
 \hline
 \ \ \ \ \ \ \ \ \ \ \ \ \ \ \ \ \ \ IoT~Devices & Active State Probability &  Mean Packet Rate &  Idle \ \ \ \ State Probability &  Mean Packet Rate\\
 \hline
Thermostat&	0.25&25&0.75&5\\
Smart Light&0.05&10&0.95&5\\
Security Camera&1&80&0&0\\
Smart Printer&0.05&75&0.95&5\\
Smart TV&0.3&120&0.7&10\\
\hline
\end{tabular}
\label{table:2}
\vspace{-5mm}
\end{table}

We finally present simulation results to evaluate the performance of the proposed IDS in a large network with many nodes, where a stealthy DDoS attack from many compromised IoT devices can actually take down a server. 

\textbf{\textit{Simulation Setup:}} 
The simulation setup consists of $10$ nodes each of which monitors $100$ devices (Fig. \ref{f:no_anomaly}). The IoT devices considered here are those that are most likely to be compromised or devices that are present in every smart home. The devices are assumed to have two states of operation, idle state and active state. The assumed probabilities of the devices being in idle or active state are given by Table \ref{table:2}. We inject the attack traffic of different rates into the simulated dataset, and apply the proposed detection and mitigation scheme to diagnose these attacks. We perform this repeatedly for different kinds of scenarios in which various combinations of devices get attacked to compute the Average Detection Delay vs. False Alarm Rate (i.e., false alarm probability) performance. 

In each node, we assume data from each IoT device is independent and identically distributed (iid) following the pattern given in Table \ref{table:2}. The probabilities listed are based on heuristics and standard day-to-day usage of the mentioned devices. For example, a smart TV is considered to be used approximately 7 to 8 hours in a day, so its active state probability is given as 0.3. 
With the shown probabilities, devices may or may not switch state after a session. We consider the following session durations: $5$ sec. for thermostat, $10$ sec. for smart light, $40$ sec. for smart printer, $900$ sec. for smart TV, and ``always on" for security camera.
The mean packet rates are determined by considering the amount of data that is transmitted per second and the average packet size. 
For each device, the number of packets are generated using the mixture of two Gaussian distributions defined by the active state probability, mean packet rates and a common standard deviation, chosen as $5$. To obtain the number of packets, the generated real-valued numbers are rounded to the nearest nonnegative integer.
For the purpose of simulations, the training data consists of 40 hours of attack-free data from 100 different devices in each node. 

\begin{figure}[t]
\centering
\includegraphics[width=.5\textwidth]{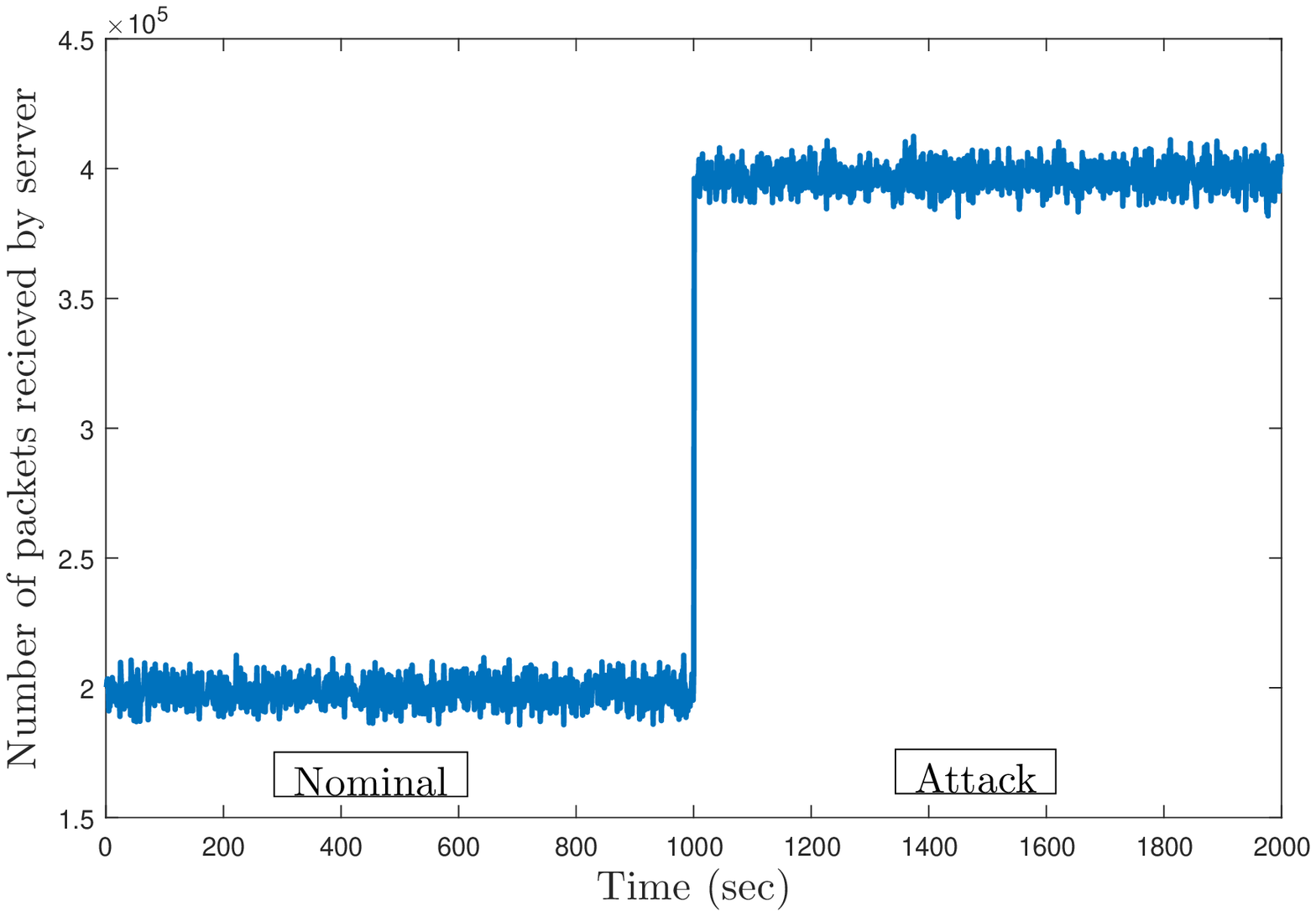}
\vspace{-3mm}
\caption{Impact of stealthy DDoS attack on the server. The attack consists of 100,000 devices with 10\% increase in their data rates.}
\label{f:tot_num-attack}
\vspace{-5mm}
\end{figure}

\textbf{\textit{Attack Model:}} Here we consider a practical scenario in which the IoT devices could be under attack, but the node is assumed to be secure. To parameterize the attack size, we consider 10\% of the devices to be compromised. The attacked devices are randomly selected to assume a general model. During the attack phase, the data rates of the selected devices are increased slightly by 10\%. To demonstrate the effectiveness of such a stealthy attack, we plotted in Fig. \ref{f:tot_num-attack} the total number of packets received by the server when attacked from 100,000 devices with 10\% increase in their data rates.

\begin{figure}[t]
\centering
\includegraphics[width=0.52\textwidth]{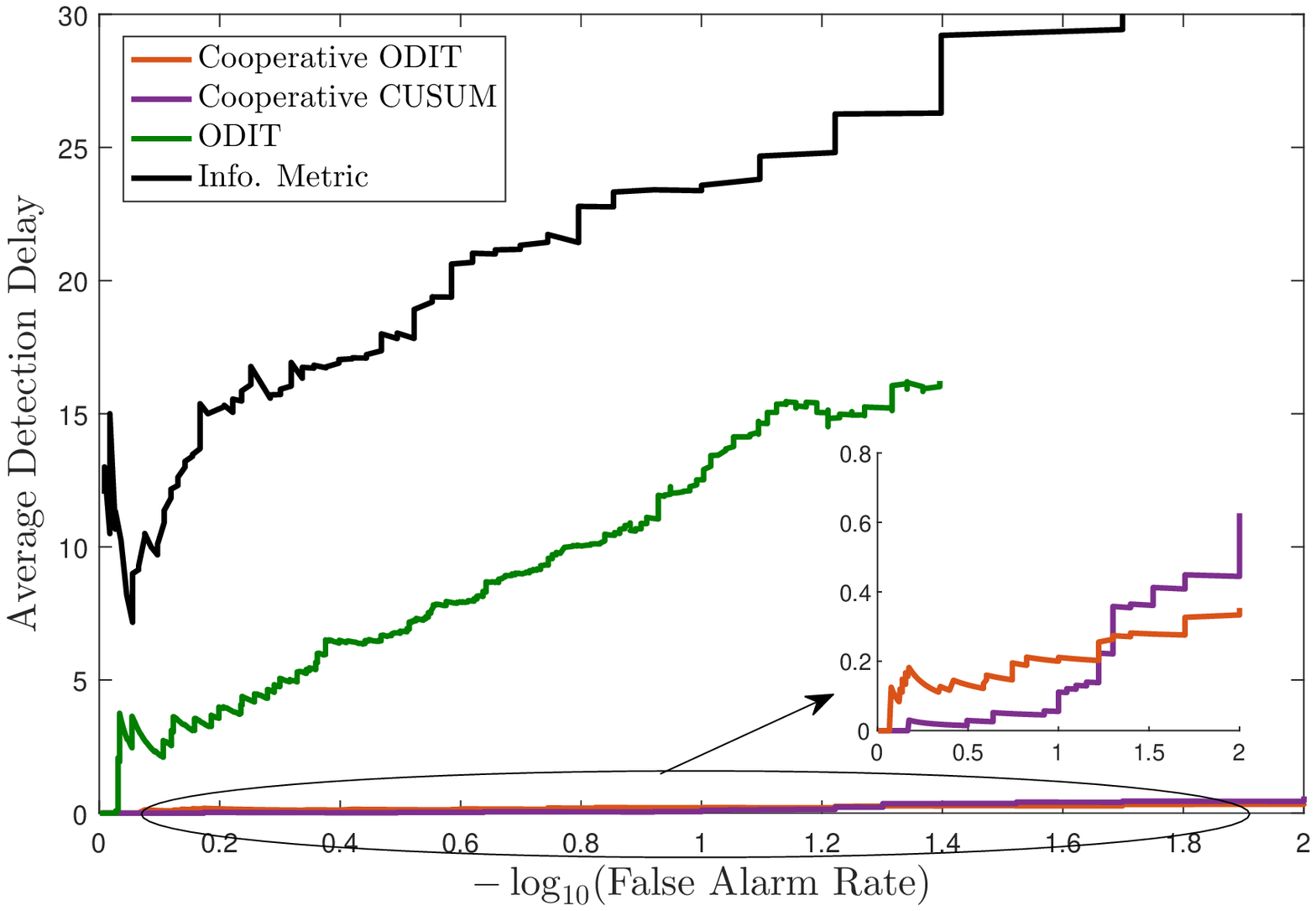}
\vspace{-3mm}
\caption{Average detection delay vs. False positive rate for the proposed cooperative ODIT-based IDS, the IDS based on cooperative CUSUM \cite{mei2010efficient}, and information metric-based IDS \cite{Informetric}. The network consists of 10 nodes each of which has 100 devices connected to it. 10\% of the devices attack with 10\% increase in data rate with respect to their nominal rates.}
\label{f:odit-mult}
\vspace{-5mm}
\end{figure}

\begin{figure}[t]
\centering
\includegraphics[width=.4\textwidth]{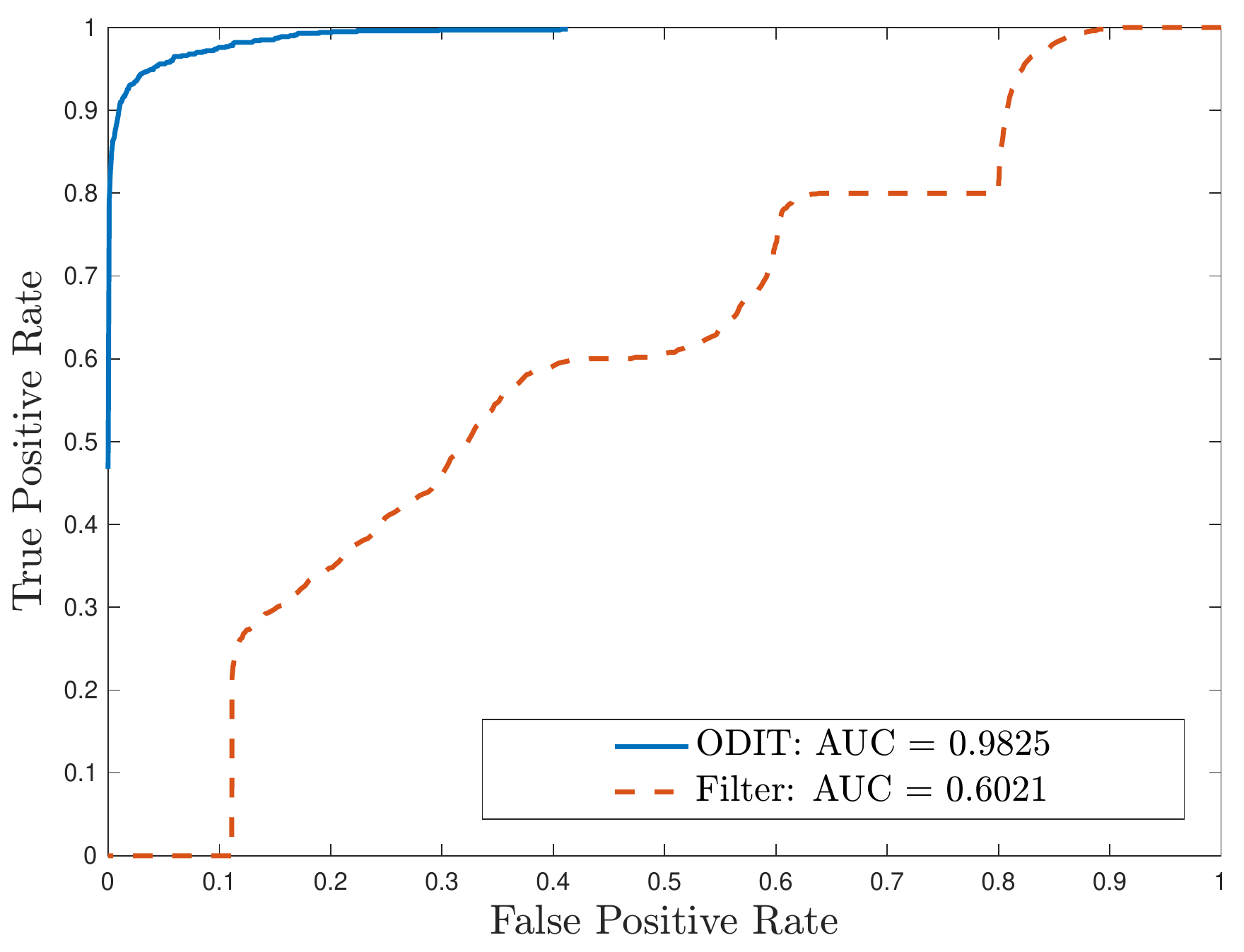}
\vspace{-3mm}
\caption{ Mitigation performance of ODIT vs. Data filtering method which applies a threshold on the data rates. After detecting an attack, the proposed IDS successfully identifies attacked devices, and blocks their traffic.}
\label{f:roc_iot}
\vspace{-5mm}
\end{figure}

\textbf{\textit{Comparisons:}} We compare our proposed model with an IDS based on cooperative CUSUM \cite{mei2010efficient}, which knows the exact parameters of the nominal model and the anomalous model. CUSUM knows exactly the mean and standard deviation of the Gaussian distribution, as well as the probability of being active for each device. Note that due to rounding to the nearest nonnegative integer value, the real probability distribution of number of packets deviates from the generative bimodal Gaussian. 
Hence, the proposed ODIT detector even sometimes outperforms CUSUM, which exactly knows the generative Gaussian model. The results for Average Detection Delay vs False Positive Rate are shown in Fig. \ref{f:odit-mult}. We see that the cooperative ODIT-based IDS, proposed in Section \ref{s:coop}, performs better than the clairvoyant CUSUM detector, which exactly knows the generative probabilistic model, for false alarm rates less than $0.1$. It significantly outperforms the information metric method proposed in \cite{Informetric}, which monitors the aggregate traffic at each node. 
In Fig. \ref{f:odit-mult}, it is seen that the cooperation among nodes facilitates earlier detection by our algorithm (ODIT vs. Cooperative ODIT). Through the proposed computationally efficient cooperation scheme, given in Section \ref{s:coop}, the ODIT-based IDS is able to handle large networks with thousands of devices. This result can be easily extended to even larger networks with millions of devices. 
Finally, in Fig. \ref{f:roc_iot}, we evaluate the mitigation performance of the proposed method (see Section \ref{sec:miti}). Since the method in \cite{Informetric} monitors the aggregate traffic at nodes, it is not straightforward for it to detect the attacking devices. Thus, to evaluate our mitigation performance, we consider the data filtering method, which simply applies a threshold to the observed raw data. The reported Area Under the Curve (AUC) values in Fig. \ref{f:roc_iot} illustrate the successful mitigation performance of the proposed method under a challenging stealthy attack scenario. 

\section{Limitations and Future Work}
\label{s:limit}

In this work, we proposed a novel intrusion detection system which is capable of quickly and accurately detecting and mitigating a broad set of IoT-empowered attacks, in particular stealthy low-rate DDoS attacks. However, there are still some limitations which need to be addressed to make the system more robust to attacks in the future. First, it is assumed that the nominal behavior of the devices does not change over time, so the IDS needs to be trained only once. However, in a real system implementation the IDS needs to be updated periodically. Secondly, feature extraction plays an important role as number of packets or packet size might not always exactly represent the characteristics of a real network. For future work, we plan to investigate other aspects of dynamic networks such as continual learning under changing nominal network traffic.

\section{Conclusion}
\label{s:conc}

With the proliferation of IoT devices, and the ease of triggering DoS attacks even by unsophisticated malicious parties, there is an increasing need for developing solutions to DDoS via IoT, especially the recent stealthy DDoS attacks. In this context, we presented a general and emerging threat model for hierarchical IoT networks. We then introduced a novel intrusion detection and mitigation framework that employs an online, scalable and nonparametric anomaly detection algorithm. Through real and simulated data, as well as an IoT testbed we evaluated the performance of proposed detection and mitigation scheme under challenging stealthy DDoS attack scenarios. Applications of the proposed scheme to large and dynamic networks with varying number of devices were also considered. 

\bibliographystyle{unsrt}
\bibliography{Ref.bib}

\vspace{-1.5cm}
\begin{IEEEbiography}
[{\includegraphics[width=1in,height=1.4in,clip,keepaspectratio]{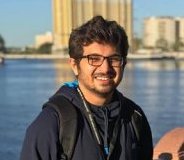}}]{Keval Doshi}
received the B.Sc. degree in Electronics and Communications Engineering from Gujarat Technological University, India, in 2017. He is currently a Ph.D. student at the Electrical Engineering Department at the University of South Florida, Tampa. His research interests include machine learning, computer vision, and cybersecurity. 
\end{IEEEbiography}

\vspace{-1.5cm}

\begin{IEEEbiography}
[{\includegraphics[width=1in,height=1.4in,clip,keepaspectratio]{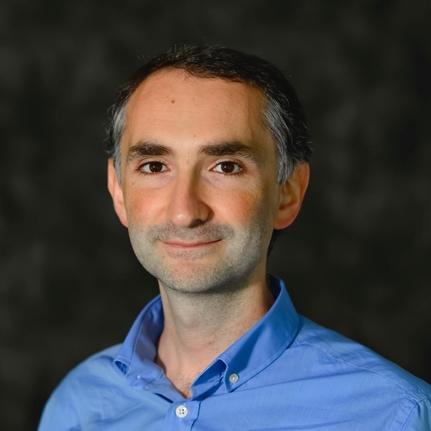}}]{Yasin Yilmaz}
(S'11-M'14) received the Ph.D. degree in Electrical Engineering from Columbia University, New York, NY, in 2014. He is currently an Assistant Professor of Electrical Engineering at the University of South Florida, Tampa. He received the Collaborative Research Award from Columbia University in 2015. His research interests include statistical signal processing, machine learning, and their applications to computer vision, cybersecurity, IoT networks, energy systems, transportation systems, and communication systems. 
\end{IEEEbiography}

\vspace{-1.5cm}

\begin{IEEEbiography}
[{\includegraphics[width=1in,height=1.4in,clip,keepaspectratio]{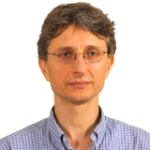}}]{Suleyman Uludag}
received the Ph.D. degree in Computer Science from DePaul University, Chicago, IL, in 2007. He is currently an Associate Professor of Computer Science at the University of Michigan - Flint. He received the Fulbright U.S. Scholar Program Core Award in 2012 and 2018. His research interests include security, privacy, and optimization of data collection particularly as applied to the Smart Grid and Intelligent Transportation Systems. 
\end{IEEEbiography}

\newpage
\title{Appendix -- Timely Detection and Mitigation of Stealthy DDoS Attacks via IoT Networks}

\author{Keval Doshi, 
Yasin Yilmaz, and Suleyman Uludag
\thanks{This work was supported in part by the U.S. National Science Foundation under the Grant CNS-1737598 and in part by the SCEEE-17-03 Grant.}
\thanks{
K. Doshi and Y.Yilmaz are with the Electrical Engineering Department, University of South Florida, Tampa, FL USA (e-mail: kevaldoshi@mail.usf.edu, yasiny@usf.edu). }
\thanks{S. Uludag, is with the Department of Computer Science, University of Michigan - Flint, MI USA (e-mail: uludag@umich.edu).}
}

\maketitle

\section*{Proof of Theorem 1}

Consider a hypersphere $\mathcal{S}_t \in \mathbb{R}^d$ centered at $\vx_t^n$ with radius $L_t^n$, the $k$NN distance of $\vx_t^n$ with respect to the training set $\cX^n_{M_2}$. The maximum likelihood estimate for the probability of a point being inside $\mathcal{S}_t$ under $f_0$ is given by $k/M_2$. It is known that, as the total number of points grow, this binomial probability estimate converges to the true probability mass in $\mathcal{S}_t$ in the mean square sense \cite{agresti2018introduction}, i.e., 
$$
k/M_2 \overset{L^2}{\to} \int_{\mathcal{S}_t} f_0(\vx) ~\text{d}\vx
$$ 
as $M_2 \to \infty$. Hence, the probability density estimate 
$$
\hat{f}_0(\vx_t^n)=\frac{k/M_2}{V_d (L_t^n)^d},
$$ 
where $V_d (L_t^n)^d$ is the volume of $\mathcal{S}_t$, converges to the actual probability density function, $\hat{f}_0(x_t^n) \overset{p}{\to} f_0(x_t^n)$ as $M_2 \to \infty$, since $\mathcal{S}_t$ shrinks and $L_t^n \to 0$. Similarly, considering a hypersphere $\mathcal{S}_{(\alpha)} \in \mathbb{R}^d$ around $\vtx_{(\alpha)}^n$ which includes $k$ points within its radius $\tilde{L}_{(\alpha)}^n$, we see that as $M_2 \to \infty$, $\tilde{L}_{(\alpha)}^n\to 0$ and 
$$
\hat{f}_0(\vtx_{(\alpha)}^n)=\frac{k/M_2}{V_d (\tilde{L}_{(\alpha)}^n)^d} \overset{p}{\to} f_0(\vtx_{(\alpha)}^n).
$$ 
Assuming a uniform distribution 
$$
f_1(\vx)=f_0(\vtx_{(\alpha)}^n), ~\forall \vx, 
$$
we conclude with
$$
\log \frac{\frac{k/M_2}{V_d (\tilde{L}_{(\alpha)}^n)^d}}{\frac{k/M_2}{V_d (L_t^n)^d}} = d \left[ \log L_t^n - \log \tilde{L}_{(\alpha)}^n \right] \overset{p}{\to} \log \frac{f_1(\vx_t^n)}{f_0(\vx_t^n)}
$$
as $M_2\to\infty$.

\section*{Proof of Theorem 2}

In online testing (see lines 6-11), the most expensive part is to compute $D_t^n$, in particular $L_t^n$. And within $L_t^n$ the expensive part is to find the $k$th nearest neighbor, which is $O(M_2 d)$ if computed straightforwardly by computing the distance of test point to all $M_2$ training points. 
The space complexity of the algorithm is due to storing $M_2$ training points, each of which is $d$-dimensional, i.e., $O(M_2 d)$.
Note that the both time and space complexity of the mitigation part shown in lines 13-23 is $O((T-\tau+1) d)$ where $T-\tau+1$ is a bounded number close to the detection delay, typically much smaller than $M_2$. 
In training, to compute $\tilde{L}_{(\alpha)}^n$ shown in line 4, $k$th nearest neighbor among $M_2$ points are computed for each of $M_1$ points, requiring $O(M_1 M_2 d)$ computations. However, training is performed once offline, so the complexity of online testing is usually critical for scalability.


\end{document}